\pdfoutput=1

\newcommand*{\ATLASLATEXPATH}{latex/}

\documentclass[cernpreprint,texlive=2016,txfonts,UKenglish]{latex/atlasdoc} 
\usepackage[subfigure=true]{\ATLASLATEXPATH atlaspackage}
\usepackage{\ATLASLATEXPATH atlasphysics}

\usepackage{multirow}
\usepackage{float}
\usepackage{lscape}

\usepackage{\ATLASLATEXPATH atlasbiblatex}
\usepackage{\ATLASLATEXPATH atlascontribute}
\usepackage{\ATLASLATEXPATH atlasphysics}

\graphicspath{{logos/}{figures/}}

\usepackage{lvqq_lvJ_paper-defs}

\AtlasTitle{Search for $WW/WZ$ resonance production in $\ell \nu qq$ final states in $pp$ collisions at $\sqrt{s} =$ 13\,\tev\ with the ATLAS detector}

\author{The ATLAS Collaboration}

\usepackage{authblk}
\makeatletter
\renewcommand\AB@affilsepx{, \protect\Affilfont}
\makeatother
 
\renewcommand\Affilfont{\itshape\small} 

\PreprintIdNumber{CERN-EP-2017-223}

\AtlasJournal{JHEP}
\AtlasJournalRef{JHEP 03 (2018) 042}
\AtlasDOI{10.1007/JHEP03(2018)042}

\AtlasAbstract{%

A search is conducted for new resonances decaying into a $WW$ or $WZ$ boson pair, where one $W$ boson decays leptonically and the other $W$ or $Z$ boson decays hadronically. It is based on proton--proton collision data with an integrated luminosity of 36.1~\ifb\  collected with the ATLAS detector at the Large Hadron Collider at a centre-of-mass energy of $\sqrt{s}$ =  13 \tev\ in 2015 and 2016. The search is sensitive to diboson resonance production via 
vector--boson fusion as well as quark--antiquark annihilation and gluon--gluon fusion mechanisms. No significant excess of events is observed with respect to the Standard Model backgrounds. Several benchmark models are used to interpret the results. Limits on the production cross section are set for a new narrow scalar resonance, a new heavy vector-boson and a spin-2 Kaluza--Klein graviton.

}

\hypersetup{pdftitle={ATLAS document},pdfauthor={The ATLAS Collaboration}}

\DeclareOldFontCommand{\rm}{\normalfont\rmfamily}{\mathrm}
\DeclareOldFontCommand{\sf}{\normalfont\sffamily}{\mathsf}
\DeclareOldFontCommand{\tt}{\normalfont\ttfamily}{\mathtt}
\DeclareOldFontCommand{\bf}{\normalfont\bfseries}{\mathbf}
\DeclareOldFontCommand{\it}{\normalfont\itshape}{\mathit}
\DeclareOldFontCommand{\sl}{\normalfont\slshape}{\@nomath\sl}
\DeclareOldFontCommand{\sc}{\normalfont\scshape}{\@nomath\sc}

\begin{document}

\maketitle

%-------------------------------------------------------------------------------
\clearpage
\section{Introduction}
\label{sec:introduction}
Diboson resonances are predicted in a number of extensions to the Standard Model (SM), such as composite Higgs models~\cite{Dugan:1984hq,Agashe:2004rs},  
warped extra dimensions~\cite{Randall:1999ee,Randall:1999vf,Davoudiasl:2000wi}, models with an extended Higgs sector~\cite{deFlorian:2016spz,Ivanov:2017dad} and grand unified theories~\cite{Pati:1974yy,Georgi:1974sy,Fritzsch:1974nn}. 
Searches for diboson resonances in various decay channels have been carried out by the ATLAS and CMS collaborations at the Large Hadron Collider (LHC), but no evidence of such resonances has been observed~\cite{EXOT-2016-01,EXOT-2013-01,CMS-EXO-13-009,CMS-EXO-13-009,EXOT-2014-18,CMS-B2G-16-007,CMS-B2G-16-004,Aaboud:2017itg,Aaboud:2017eta}. The most recent ATLAS searches using data collected in 2015 and 2016 have been performed in the $ZZ/ZW$ final state~\cite{Aaboud:2017itg}, with one $Z$ decaying to leptons, and the fully hadronic final state with boson-tagged jets~\cite{Aaboud:2017eta}.

This paper reports on a search for a charged or neutral resonance, in a mass range from 300\,\gev\ to 5000\,\gev, that decays into a $WZ$ or $WW$ boson pair. The semileptonic final state where one $W$ boson decays leptonically ($W \to \ell \nu$ with  $\ell = e, \mu$) and the other $W/Z$ boson (denoted by $V$) decays hadronically ($V \to q\bar{q}^\prime/q\bar{q}$ with $q,q^\prime$ quarks) is considered. The search uses $pp$ collision data at a centre-of-mass energy of 13 \tev, corresponding to an integrated luminosity of 36.1 \ifb, collected  by the ATLAS experiment in 2015 and 2016. The strategy for identification of resonances depends on the ability to resolve the quarks from the hadronically decaying $V$ boson. For high-mass resonances, the opening angles between the quarks from $V$ boson decays are small and both quarks can be identified as a single jet. This case is referred to as the merged analysis and is denoted by $\ell\nu J$. In contrast, separate identification of the two quarks from low-mass resonances is referred to as the resolved analysis and is denoted by $\ell\nu jj$. 

In addition to a larger data sample, the search makes use of several improvements to the methodology compared to the previous ATLAS result~\cite{EXOT-2016-01}. The resolved analysis has been included,  and in addition the event selections are optimized for two different production modes: the vector-boson fusion (VBF) and the gluon--gluon fusion (ggF) or quark--antiquark ($\mathrm{q \bar{q}}$) annihilation. In addition, a new mass reconstruction algorithm is implemented for hadronically decaying $W/Z$ bosons that are highly Lorentz boosted. It is based on both the calorimeter energy deposits and the charged tracks instead of calorimeter information alone, as used in the previous publication~\cite{EXOT-2016-01}.

The VBF process ($pp \to VVjj$) is characterized by the presence of two jets with a large rapidity gap resulting from quarks from which a vector boson is radiated. The absence of this topology is interpreted as ggF or $\mathrm{q\bar{q}}$ production, collectively referred to as $\mathrm{ggF/q\bar{q}}$ in this paper. Results are provided for the VBF and $\mathrm{ggF/q\bar{q}}$ categories separately and possible signal leakage between categories is neglected.

The spectrum of the reconstructed invariant mass of the $WV$ resonance candidates, $m(WV)$, is examined for localized excesses over the expected SM background. Three signal models are used to optimize the event selection, assess the sensitivity of the search and interpret the data: an additional heavy Higgs boson predicted by many theories beyond the SM, a heavy vector triplet (HVT) parameterization based on a simplified phenomenological Lagrangian~\cite{Pappadopulo:2014qza,deBlas:2012qp} and a bulk Randall--Sundrum (RS) model~\cite{Agashe:2007zd}.

%-------------------------------------------------------------------------------

\section{ATLAS detector}
\label{sec:detector}
The ATLAS detector~\cite{PERF-2007-01} is a general-purpose particle detector used to investigate a broad range of physics processes. It includes an inner detector (ID) surrounded by a superconducting solenoid, electromagnetic (EM) and hadronic calorimeters and a muon spectrometer (MS) inside a system of toroidal magnets. The ID consists of a silicon pixel detector including a newly installed innermost layer called the insertable B-layer~\cite{IBL}, a silicon microstrip detector and a straw-tube tracker. It is immersed in a $2$ T axial magnetic field and provides precision tracking of charged particles with pseudorapidity\footnote{ATLAS uses a right-handed coordinate system with its origin at the nominal interaction point (IP) in the centre of the detector and the $z$-axis along the beam pipe. The $x$-axis points from the IP to the centre of the LHC ring, and the $y$-axis points upwards. Cylindrical coordinates $(r,\phi)$ are used in the transverse plane, $\phi$ being the azimuthal angle around the beam pipe. The pseudorapidity is defined in terms of the polar angle $\theta$ as $\eta = -\ln \tan(\theta/2)$. Angular distance is measured in units of $\Delta R \equiv \sqrt{(\Delta\eta)^{2} + (\Delta\phi)^{2}}$.} $|\eta| < 2.5$. The straw-tube tracker also provides transition radiation measurements for electron identification. The calorimeter system comprises finely segmented sampling calorimeters using lead/liquid-argon for the detection of EM showers up to $|\eta| =$ 3.2, and (copper or tungsten)/liquid-argon for hadronic showers for 1.5 $< |\eta| <$ 4.9. In the central region ($|\eta| <$ 1.7), a steel/scintillator hadronic calorimeter is used. Outside the calorimeters, the muon system incorporates multiple layers of trigger and tracking chambers in a magnetic field produced by a system of superconducting toroids, enabling an independent precise measurement of muon track momenta for $|\eta| <$ 2.7. The ATLAS detector has a two-level trigger system that is based on custom hardware followed by a software trigger to reduce the selected event rate to approximately 1 kHz for offline analysis~\cite{TRIG-2016-01}.

%-------------------------------------------------------------------------------

\section{Signal and background simulation}
\label{sec:simul}
Samples of simulated signal and background events are used to optimize the event selection and to estimate the background contribution from various SM processes. For all the signal samples, the $\ell \nu qq$ final state is imposed at the generator level.

The heavy neutral Higgs boson signal was generated using \textsc{Powheg-Box} v1~\cite{Frixione:2007vw,Alioli:2010xd} with the next-to-leading-order (NLO) gg\_H~\cite{Alioli:2008tz} and VBF\_H~\cite{Nason:2009ai} modules and the CT10~\cite{Lai:2010vv} parton density functions (PDF). The \textsc{Powheg-Box} event generator was interfaced to \textsc{Pythia} 8.186~\cite{Sjostrand:2007gs} for parton showering, underlying event and hadronization using the AZNLO set of tuned parameters (tune)~\cite{STDM-2012-23} and the CTEQ6L1 PDF~\cite{Pumplin:2002vw}. Possible interference effects with the SM diboson production were neglected. Scalar $WW$ resonances with masses ranging from 300 \gev\ to 3000 \gev\ were generated with a narrow width; they were produced via either the ggF or VBF process~\cite{Hill:1987ea,Barger:2007im}.

For interpretation in terms of a vector resonance produced via $q \bar{q}$ annihilation, simulated $Z' \to WW$ and $W' \to WZ$ samples of two benchmark models based on the HVT $q \bar{q}$ parameterized Lagrangian~\cite{Pappadopulo:2014qza,deBlas:2012qp} were generated. Model A, with a strength of the vector-boson interaction $g_V=1$, is typical of an extended gauge model~\cite{Barger:1980ix} with the heavy vectors having comparable branching ratios into fermions and gauge bosons. Model B, with $g_V = 3$, is representative of composite Higgs models, where the fermionic couplings are suppressed~\cite{Contino:2011np,Bellazzini:2014yua,Panico:2015jxa}. In both scenarios, the resonance width is narrower than the detector resolution. The HVT VBF Model samples were generated with the coupling to fermions set to zero, and the couplings to gauge bosons similar to those of Model A. The signal samples were produced in a mass range from 300\,\gev\,to 5000\,\gev\ using \textsc{MadGraph5\_aMC@NLO} 2.2.2~\cite{Alwall:2014hca} interfaced to \textsc{Pythia} 8.186 with the NNPDF23\_lo~\cite{Ball:2012cx} PDF and the A14 tune~\cite{ATL-PHYS-PUB-2014-021} for the underlying event.

The $\mathrm{KK}$ graviton ($G_\mathrm{KK}\to WW$)~\cite{Agashe:2007zd} signal produced via ggF with masses from 300\,\gev\,to 5000\,\gev\, was generated using \textsc{MadGraph5\_aMC@NLO} with a model given in Ref.~\cite{deAquino:2011ix}. The $G_\mathrm{KK}$ is the first Kaluza--Klein mode~\cite{KKmodesHLZ} of a spin-2 graviton in a warped extra dimension with curvature $k$ and dimensionless coupling constant $k/\bar{M}_{{\text{Pl}}}\sim\mathcal{O}(1)$, where $\bar{M}_{\text{Pl}}$ is the reduced Planck mass. A bulk RS $G_\mathrm{KK}$ with $k/\bar{M}_{{\text{Pl}}} =$ 1.0 is considered in this paper.
%For the interpretation of the results, a bulk RS $G_\mathrm{KK}$ with $k/\bar{M}_{{\text{Pl}}} =$ 1.0 is considered.

The dominant SM background arises from events with a $W$ boson produced in association with jets ($W$+jets). Additional sources of SM background include the production of top quarks, multijets, dibosons and $Z$+jets. Events containing $W$ or $Z$ bosons with associated jets were simulated using the \textsc{Sherpa} 2.2.1~\cite{Gleisberg:2008ta} generator with the NNPDF30\_nnlo~\cite{Ball:2014uwa} PDF. Matrix elements were calculated for up to 2 partons at NLO and 4 partons at LO using Comix~\cite{Gleisberg:2008fv} and OpenLoops~\cite{Cascioli:2011va} and merged with the Sherpa parton shower~\cite{Schumann:2007mg} using the ME+PS@NLO prescription~\cite{Hoeche:2012yf}. To estimate systematic uncertainties related to the $V$+jets processes, alternative samples were generated using \textsc{MadGraph5\_aMC@NLO} 2.2.2 interfaced to the \textsc{Pythia} 8.186 parton shower model, using the A14 tune together with the NNPDF23\_lo PDF. For the generation of $t\bar{t}$ and single top quarks the \textsc{Powheg-Box} 2~\cite{Nason:2004rx} generator with the CT10 PDF in the matrix element calculations was used. Systematic uncertainties associated with showering and hadronization are evaluated using alternative \textsc{Powheg-Box} samples interfaced with \textsc{Herwig++} 2.7.1~\cite{Bahr:2008pv} and using the UEEE5 underlying event tune~\cite{Gieseke:2012ft}. Additional systematic uncertainties related to the shape of the $WV$ mass are computed using alternative samples generated by \textsc{MadGraph5\_aMC@NLO} 2.2.2~\cite{Alwall:2014hca} with the CT10 PDF. Diboson samples ($WW$, $WZ$ and $ZZ$) were generated using \textsc{Sherpa}\,2.2.1 with the CT10 PDF. Additional diboson events using the \textsc{Powheg-Box} generator, interfaced to the \textsc{Pythia} 8.186 parton shower model, were generated for the purpose of estimating systematic uncertainties. The CT10NLO set was used for the PDF of the hard-scatter process and the CTEQ6L1 PDF was used for the parton shower. All the background cross sections were computed to the next-to-next-leading order (NNLO) in QCD~\cite{Anastasiou:2003ds,Czakon:2013goa,Czakon:2011xx,Kidonakis:2010tc,Kidonakis:2010ux}, except for the diboson samples for which the generator cross sections at NLO are used. \textsc{EvtGen} 1.2.0~\cite{Lange:2001uf} was used for simulating the bottom and charm hadron decays, except for samples generated by \textsc{Sherpa}. The multijet background estimation relies purely on data-driven techniques.

The effect of multiple $pp$ interactions in the same and neighboring bunch crossings (pile-up) was included by overlaying minimum-bias events simulated with {\textsc Pythia} 8.186 on each generated signal and background event. The number of overlaid events was reweighted in such a way that the distribution of the average number of interactions per $pp$ bunch crossing in the simulation matches that observed in the data. The generated samples were processed through a {\textsc GEANT4}-based detector simulation~\cite{Agostinelli:2002hh,SOFT-2010-01} and the standard ATLAS reconstruction software.

%-------------------------------------------------------------------------------
\section{Event reconstruction}
\label{sec:objects}
Events are required to have at least one primary vertex with at least two associated tracks, each with transverse momentum \pT > 0.4 \gev. If there is more than one primary vertex reconstructed in the event, the one with the largest track $\sum\pT^2$ is chosen as the hard-scatter primary vertex and is subsequently used for the reconstruction of electrons, muons, jets and missing transverse momentum. Only events with exactly one ``signal'' lepton and no additional ``veto'' leptons, as defined later in this section, are selected.

Electrons are reconstructed from clusters of energy deposits in the EM calorimeter that match a track reconstructed in the ID. They are identified using a likelihood identification criterion described in Ref.~\cite{ATLAS-CONF-2016-024}. ``Signal'' electrons must satisfy ``tight'' identification criteria and have transverse momentum \pT > 27\,\gev, while ``veto'' electrons are required to pass the ``loose'' identification criteria and \pT > 7\,\gev. All electrons have to satisfy $|\eta| <$ 2.47, excluding the transition region between the barrel and endcaps (1.37 < $|\eta|$ < 1.52). Electron candidates are further required to be isolated from other tracks and energy depositions in the calorimeter. This is achieved by examining the scalar sum of transverse momenta of tracks and the sum of transverse energy deposits~\cite{ATL-PHYS-PUB-2016-015} within a cone of size $\Delta R =$ 0.2 around the electron, excluding the transverse energy of the electron itself and correcting for the expected pile-up contributions. The isolation requirement for electrons is chosen to ensure approximately 95\% and 99\% selection efficiency, for signal and veto electrons, respectively.

Muons are reconstructed by combining an ID track with an MS track that has compatible trajectory~\cite{PERF-2015-10}. Based on the quality
of their reconstruction and identification, signal muons are required
to pass the ``medium'' selection with \pt > 27\,\gev\, and $|\eta|$ < 2.5,
while veto muons are required to pass the ``loose'' selection, \pt > 7\,\gev\,and $|\eta|$ < 2.7. In addition, a similar isolation requirement to that used for electron candidates, only using tracks within a cone of $\Delta R=0.3$, is applied to signal and veto muon candidates with an efficiency of 99\%.

To reject non-prompt leptons, requirements of $|d_0|/\sigma_{d_0}<5\,(3)$ and $|z_0 \sin\theta|<0.5$~mm are imposed on the tracks associated with the electrons (muons), where $d_0$ is the transverse impact parameter with respect to the measured beam-line position, $\sigma_{d_0}$ is the corresponding uncertainty, $z_0$ is the longitudinal impact parameter with respect to the primary vertex and $\theta$ is the polar angle of the track.\footnote{The transverse impact parameter, longitudinal impact parameter and polar angle are calculated at the point of closest approach of the track to the beam line.}

Jets are reconstructed  using the anti-$k_t$ algorithm~\cite{antikt} implemented in the FastJet package~\cite{Cacciari:2011ma} from three-dimensional topological clusters of energy deposits in the calorimeter~\cite{Aad:2016upy}, with two different radius parameters: $R = 1.0$ for large-$R$ jets (denoted by $J$) and $R = 0.4$ for small-$R$ jets.

Small-$R$ jets~\cite{PERF-2016-04} are required to have \pt > 20 \gev\,  and $|\eta|$ < 2.4, while jets considered for the tagging of VBF events are required to have \pt > 30 \gev\, and $|\eta|$ < 4.5. For jets with \pt < 60 \gev\, and $|\eta| < 2.4$ a jet-vertex-tagger multivariate discriminant~\cite{Aad:2015ina}, based on tracking and vertexing information, is applied to select jets that originate from the primary vertex. The selected working point provides at least 92\% efficiency.

An overlap removal procedure is applied to prevent using the same energy deposits in more than one electron, muon or jet. Small-$R$ jets are discarded if they are within a cone of size $\Delta R = 0.2$ around the direction of an electron candidate. However, if the distance between a jet and an electron candidate is within $0.2 < \Delta R < \min(0.4, 0.04 + 10/\pt^e)$, the jet is retained but the nearby electron is rejected from the analysis. A muon candidate lying within $\Delta R <  \min(0.4, 0.04 + 10/\pt^\mu)$ from a small-$R$ jet is discarded unless it is within $\Delta R < 0.2$ and 
satisfies one of the two following requirements: 
(a) the small-$R$ jet has fewer than three tracks;
(b) $\pt^\mu/\pt^j > 0.5$ and $\pt^\mu/\sum{\pt} > 0.7$, where $\sum{\pt}$ is the sum of the transverse momenta of tracks associated with the small-$R$ jet. In this case, the muon is retained but the nearby small-$R$ jet is rejected.

Small-$R$ jets containing $b$-hadrons are identified using the MV2c10 $b$-tagging algorithm~\cite{PERF-2012-04,ATL-PHYS-PUB-2016-012} with an efficiency of 85\%, determined with $\ttbar$ simulated events. The corresponding misidentification rates are approximately  3\% and 30\% for selecting jets originating from light quark and charm quark, respectively. For simulated samples the $b$-tagging efficiencies are corrected to match those measured in data~\cite{PERF-2012-04}.

Large-$R$ jets~\cite{ATL-PHYS-PUB-2015-033,PERF-2015-03} are formed from constituent energy deposits and are trimmed to mitigate pile-up effects and soft radiation. The jet constituents are reclustered into subjets using the $k_t$ algorithm with $R = 0.2$~\cite{Catani:1993hr}, removing those which carry less than 5\%\ of the \pt\ of the original jet~\cite{Krohn:2009th}. To overcome the limited angular resolution of the calorimeter, the mass of a large-$R$ jet is computed using a combination of calorimeter and tracking information~\cite{conf:2016-035}. The mass is defined as:

\[
m_{J} \equiv w_{\rm calo} \times m_J^\mathrm{calo} + w_{\rm track} \times \Big( m_J^\mathrm{track} \frac{\pt^\mathrm{calo}}{ \pt^\mathrm{track}} \Big),
\]
where $m_J^\mathrm{track}$ ($m_J^\mathrm{calo}$) and $\pt^\mathrm{track}$ ($\pt^\mathrm{calo}$) are the invariant mass and total transverse momentum obtained when using the tracks (calorimeter energy clusters) associated with the large-$R$ jet, respectively. In this study, charged tracks with $\pt > 0.4\,\gev$ are matched to large-$R$ jets using ghost association~\cite{Cacciari:2007fd}. To correct for the missing neutral component in the calculation of the track-based jet mass,  $m_J^\mathrm{track}$ is scaled by the ratio of calorimeter to track \pt estimates. The weighting factors $w_{\rm calo}$ and $w_{\rm track}$, with $w_{\rm calo} + w_{\rm track} = 1$, are \pt-dependent functions of the calorimeter and track-based jet mass resolutions which optimize the combined jet mass resolution. Large-$R$ jets are required to have $\pt > 200\,\gev$, $|\eta|<2.0$, $m_J>50\,\gev$ and an angular separation of $\Delta R > 1.0$ from signal electrons.
%In addition, the measured large-$R$ jet momentum is recalculated as $p = p^\mathrm{calo} \times (m_J/m^\mathrm{calo}_J)$. 

A jet substructure variable, $D_2$~\cite{Larkoski:2014gra}, is used to classify large-$R$ jets. The $D_2$ variable\footnote{The angular exponent $\beta$, defined in Ref.~\cite{Larkoski:2014gra}, is set to unity.} is defined as a ratio of two- and three-point energy correlation functions~\cite{Larkoski:2014gra,Larkoski:2015kga}, which are based on the energies and pairwise angular distances of particles within a jet. This variable is optimized~\cite{conf:2017-064} to distinguish between jets originating from a single parton and those coming from the two-body decay of a heavy particle. 

In the merged analysis, a baseline selection on the $D_2$ variable providing 80\% efficiency for $V$ signals is applied to all large-$R$ jets. To further distinguish hadronically decaying $V$ bosons from jets originating from non-top quarks or gluons, boson tagging algorithms ($V$-tagging) based on the combined large-$R$ jet mass and the $D_2$ variable are constructed. The requirements on the $D_2$ variable and the mass window depend on the jet \pt\ and are defined separately for the $W$ and $Z$ bosons. In this paper, working points resulting in 50\%\ and 80\%\ signal selection efficiency are used, as defined in Sec.~\ref{sec:selection}.

The missing transverse momentum (\met) is the absolute value of the negative vectorial sum of the transverse momenta of electrons, muons, and small-$R$ jets. Reconstructed charged-particle tracks originating from the primary vertex and not matched to any electron, muon, or jet are also included in the \met\ reconstruction~\cite{ATL-PHYS-PUB-2015-023}.

The neutrino momentum four-vector is reconstructed by imposing a $W$ boson mass constraint on the charged-lepton--neutrino system. The neutrino transverse momentum components are set equal to the missing transverse momentum of the event and the unknown $z$-component of the momentum ($p_z$) is obtained from the resulting quadratic equation. The $p_z$ is chosen as either the smaller, in absolute value, of the two real solutions or, if the solution is complex, its real part. 

The selection criteria outlined above are the result of a signature-dependent optimisation using the asymptotic significance.

%-------------------------------------------------------------------------------
\section{Trigger and event selection}
\label{sec:selection}
\label{sec:selection}

Events are selected that contain exactly one charged signal lepton and no additional veto electrons or muons. Single-electron triggers with minimum transverse energy (\et) thresholds of 24 \gev\ and 26 \gev\ in 2015 and 2016, as well as 60 \gev\ are applied to record events in the electron final state. The low threshold triggers require electron candidates to pass isolation requirements resulting in at least 90\% efficiency, depending on the lepton \pt. As for the muon final state, the events are recorded either by a single-muon trigger or an \met\ triggers. The single-muon trigger, with \pt > 20 (26) \gev\ in 2015 (2016), is subject to a large inefficiency due to limited trigger hardware coverage. The \met\ trigger has  an online threshold of 70\,\gev\ for the 2015 data and of 90--110 \gev\ for the 2016 data, where the muon track \pt\  is not used to compute \met\  in the trigger algorithm. Therefore, it is fully efficient for $W \to \mu \nu$ with  \pt($W$)  > 200 \gev\ and it is used in the merged analysis, where a high-\pt lepton is expected, to recover the single-muon trigger inefficiency. Events recorded by single-lepton triggers, where the signal lepton matches the trigger lepton, and \met\ triggers are selected.

The sensitivity to resonances of different masses is optimized by classifying the events according to the topology, production mechanism and amount of background. The event selection criteria are summarized in Tables~\ref{table:sel_merged} and~\ref{table:sel_resolved} for the merged and resolved analyses respectively. Figure~\ref{fig:reg_schem} illustrates the jet selections used to reconstruct the hadronically decaying $V$ boson candidates in the signal and control regions of the analysis. The mass of either the large-$R$ jet ($m_J$) or the system of two small-$R$ jets ($m_{jj}$) is used to define ``mass windows''. 

\begin{figure}[h]
\centering
\subfigure[][\label{fig:MergedDef}]{\includegraphics[width=0.48\linewidth]{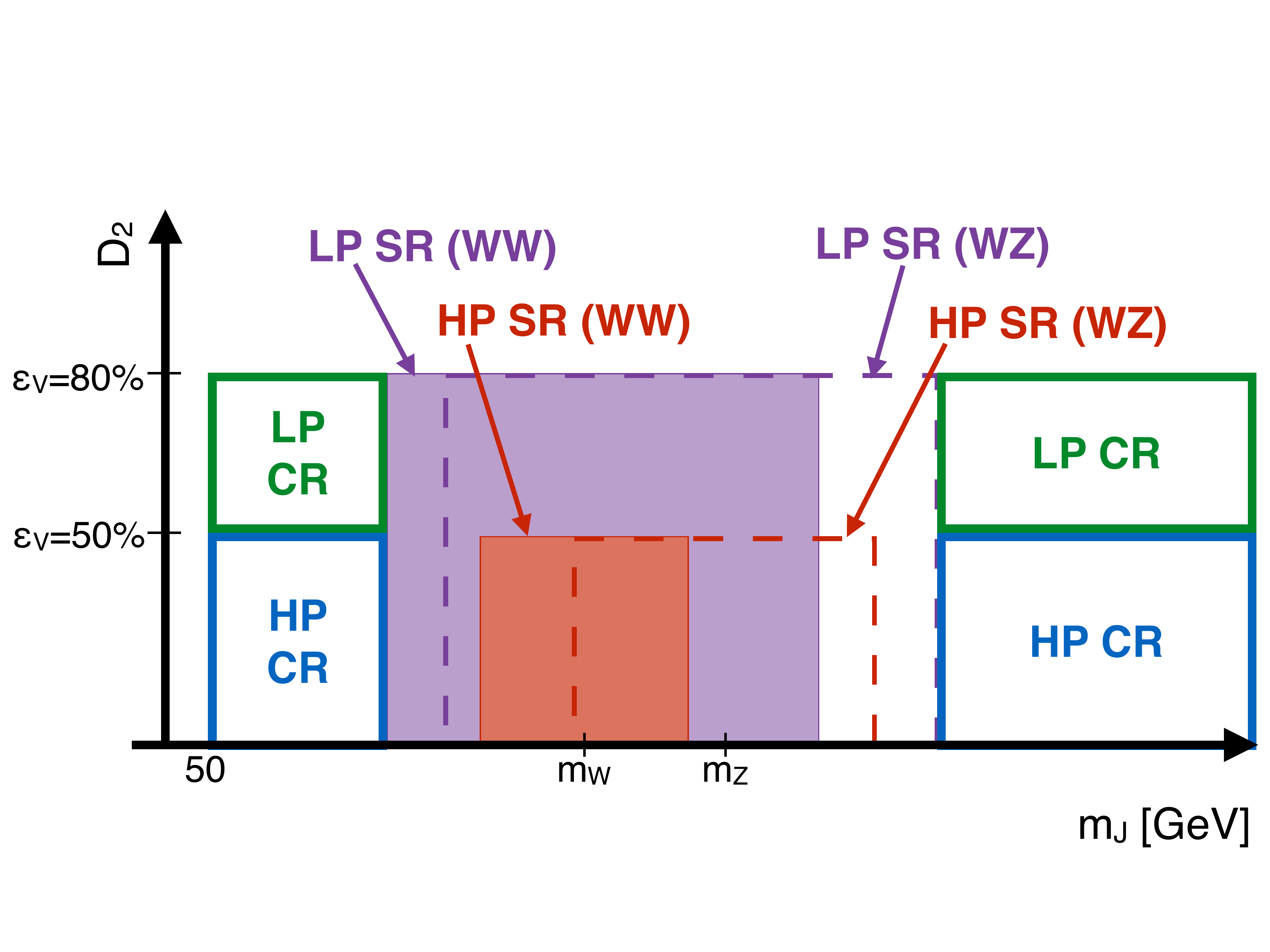}}
\subfigure[][\label{fig:ResolvedDef}]{\includegraphics[width=0.48\linewidth]{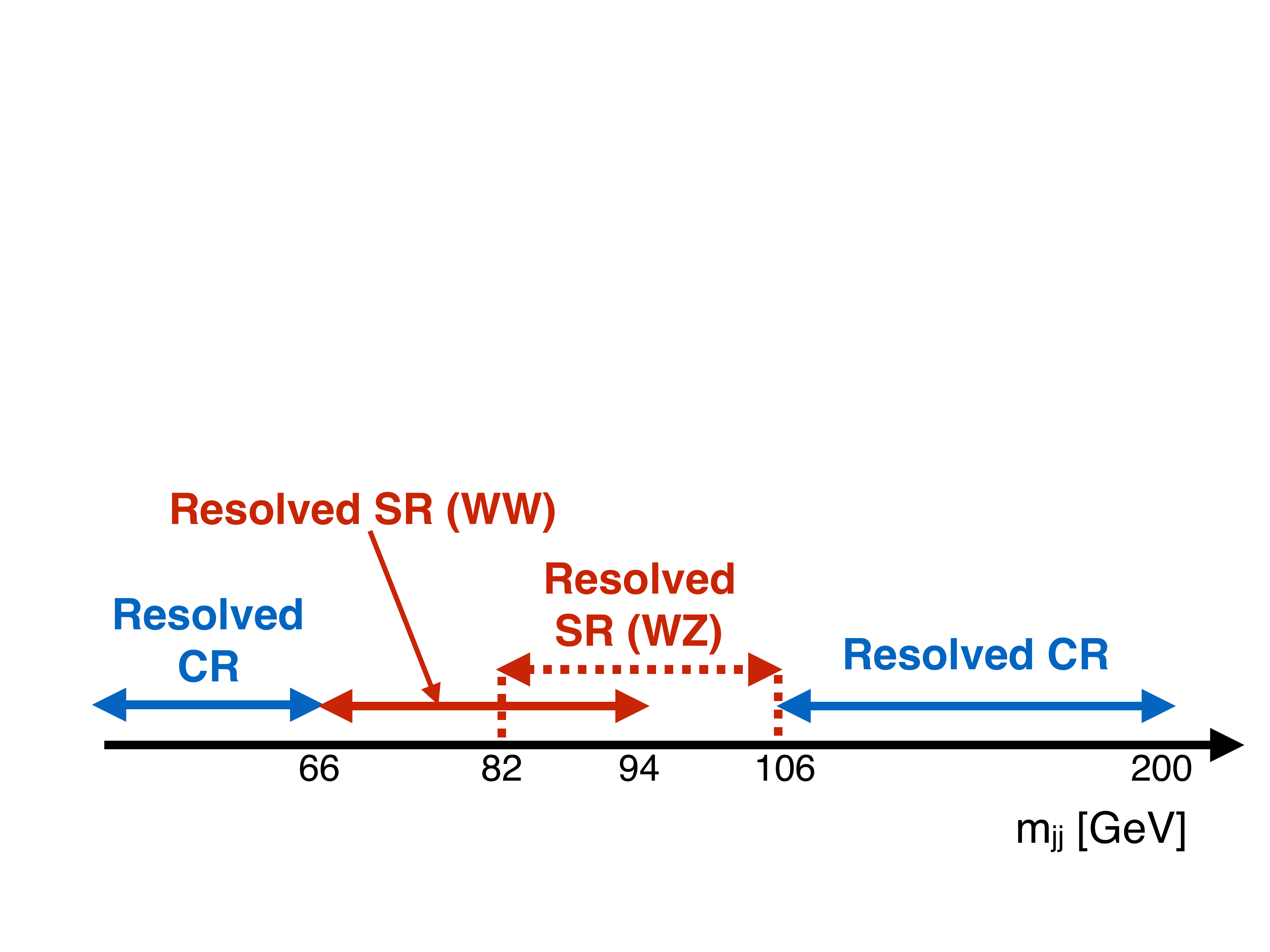}}
\caption{\subref{fig:MergedDef} Illustration of the merged $WW$ (shaded area) and $WZ$ (dashed lines) signal regions (SR) according to the large-$R$ jets selection. The 50\% and 80\% $V$-tagging efficiency ($\varepsilon_V$) working points, based on the combined cut of the $D_2$ and $m_J$, are used to form the high-purity (HP) and low-purity (LP) regions respectively. For each working point, a jet mass requirement is imposed and an upper bound on the substructure variable is set. Since both requirements depend on the \pt\ of the large-$R$ jet, an absolute definition is not given in the figure. \subref{fig:ResolvedDef} Definitions of the resolved $WW$ and $WZ$ SR based on the dijet mass selection. In both channels, the SR mass sidebands are used to define the $W$+jets control region (CR). 
}
\label{fig:reg_schem}
\end{figure}

\begin{table}[h]
\centering
\caption{Summary of the selection criteria used to define the merged $WW$ and $WZ$ signal regions (SR) and their corresponding $W$+jets control regions ($W$ CR) and $t\bar{t}$ control regions ($t\bar{t}$ CR) in the high-purity (HP) and low-purity (LP) categories. The events are also categorized according to their production mechanism, the VBF selection is prioritized and the remaining events are assigned to the $\mathrm{ggF/q\bar{q}}$ category.}
\vspace{5pt}
\label{table:sel_merged}
\begin{tabular}{|l|l|c|c|c|}
\hline
\multicolumn{2}{|l|}{Selection} & SR: HP (LP)  &  $W$ CR: HP (LP)  & $t\bar{t}$ CR: HP (LP) \\
\hline
\multirow{2}{*}{Production category}  & VBF& \multicolumn{3}{c|}{ $m^{\rm tag}(j,j) >$ 770\,\gev\ and $|\Delta \eta^{\rm tag} (j,j)| >4.7$ }\\
\cline{2-5}
 & $\mathrm{ggF/q\bar{q}}$ & \multicolumn{3}{c|}{          Fails VBF selection}\\
\hline
\multirow{4}{*}{$W\rightarrow \ell\nu$ selection} & Num. of signal leptons & \multicolumn{3}{c|}{ 1 } \\
\cline{2-5}
&Num. of veto leptons & \multicolumn{3}{c|}{ 0 }  \\
\cline{2-5}
&\vphantom{\Large B} \met & \multicolumn{3}{c|}{ $>$ 100\,\GeV } \\
\cline{2-5}
&$\pt(\ell\nu)$ & \multicolumn{3}{c|}{ $>$ 200\,\GeV } \\
\cline{2-5}
&$\met / \pt(e \nu)$ & \multicolumn{3}{c|}{$>$ 0.2 } \\
\hline
\multirow{4}{*}{$V \rightarrow J$ selection} & Num. of large-$R$ jets & \multicolumn{3}{c|}{ $\geq 1$ } \\
\cline{2-5}
& \vphantom{\Large B}$D_2$ eff. working point (\%) & Pass 50 (80)  &  Pass 50 (80)&  Pass 50 (80)\\   
\cline{2-5}         
& \vphantom{\Large B}Mass window                         &               &              &             \\
& \vphantom{\Large B}Eff. working point (\%) & Pass 50 (80)  &  Fail 80 (80)& Pass 50 (80)\\
\hline
\multirow{2}{*}{Topology criteria}
& $\pt(\ell\nu) / m(WV)$ & \multicolumn{3}{c|}{ \multirow{2}{*}{$>$ 0.3 for VBF and $>$ 0.4 for $\mathrm{ggF/q\bar{q}}$ category } }\\
& $\pt(J) / m(WV)$  & \multicolumn{3}{c|}{} \\
\hline
\multirow{2}{*}{Num. of $b$-tagged jet} & excluding $b$-tagged jets with & \multicolumn{2}{c|}{\multirow{2}{*}{0}} & \multirow{2}{*}{$\geq 1$} \\
                                  & $\Delta R(J,b) \leq$ 1.0& \multicolumn{2}{c|}{}                    &                            \\ 
\hline
\end{tabular}
\end{table}

\begin{table}[h]
  \caption{Summary of the selection criteria in the resolved analysis for the $WW$ and $WZ$ signal regions (SR), $W$+jets control region ($W$ CR) and $ t\bar{t}$ control region ($t\bar{t}$ CR). The events are also categorized according to their production mechanism, the VBF selection is prioritized and the remaining events are assigned to the $\mathrm{ggF/q\bar{q}}$ category.}
\begin{center} 
\label{table:sel_resolved}
\begin{tabular}{|l|l|c|c|c|}
\hline
\multicolumn{2}{|l|}{Selection} & $WW$ ($WZ$) SR  & $W$ CR & $t\bar{t}$ CR \\
\hline   
\multirow{2}{*}{Production category} & VBF& \multicolumn{3}{c|}{ $m^{\rm tag}(j,j) >$ 770\,\gev\ and $|\Delta \eta^{\rm tag} (j,j)| >4.7$}\\
\cline{2-5}
& $\mathrm{ggF/q\bar{q}}$ & \multicolumn{3}{c|}{          Fails VBF selection}\\
\hline
\multirow{4}{*}{$W\rightarrow \ell\nu$ selection} & Num. of signal leptons & \multicolumn{3}{c|}{ 1 } \\
\cline{2-5}
&Num. of veto leptons & \multicolumn{3}{c|}{ 0 }  \\
\cline{2-5}
&\met & \multicolumn{3}{c|}{ $>$ 60\,\GeV } \\
\cline{2-5}
&$\pt(\ell\nu)$ & \multicolumn{3}{c|}{ $>$ 75\,\GeV } \\
\cline{2-5}
&$\met / \pt(e \nu)$ & \multicolumn{3}{c|}{$>$ 0.2 } \\
\hline
\multirow{3}{*}{$V \rightarrow j_1j_2$ selection} & Num. of small-$R$ jets & \multicolumn{3}{c|}{ $\geq2$ } \\ 
\cline{2-5}
& $\pt (j_1)$ & \multicolumn{3}{c|}{ $>$ 60 \GeV}\\
\cline{2-5}
& $\pt (j_2)$ & \multicolumn{3}{c|}{ $>$ 45 \GeV}\\
\cline{2-5}
& $m(j_1j_2)$ [\,\gev] & [66, 94] &    $<$ 66 &  [66, 106] \\
 &  & ([82, 106])  &  or [106, 200] & \\
\hline   
\multirow{5}{*}{Topology criteria} & $\Delta\phi(j,\ell)$ & \multicolumn{3}{c|}{ $>$ 1.0}\\
\cline{2-5}
& $\Delta\phi(j,\met)$ & \multicolumn{3}{c|}{ $>$ 1.0}\\
\cline{2-5}
& $\Delta\phi(j,j)$ & \multicolumn{3}{c|}{ $<$ 1.5}\\
\cline{2-5}
& $\Delta\phi(\ell,\met)$ & \multicolumn{3}{c|}{ $<1.5$}\\
\cline{2-5}
& $\pt(\ell\nu) / m(WV)$ &\multicolumn{3}{c|}{\multirow{2}{*}{$>$ 0.3 for VBF and 0.35 for $\mathrm{ggF/q\bar{q}}$ category}}\\
& $\pt(j_1j_2) / m(WV)$ & \multicolumn{3}{c|}{ } \\
\hline
\multirow{4}{*}{Num. of $b$-tagged jets} &  $j_1 \equiv b$ or $j_2 \equiv b$  &              &          & $>$ 0 \\
                                  & where $V \to j_1 j_2$     & $\leq 1 (2)$ & $\leq 1$ & (for jets other \\  
\cline{2-4}
& $j_1 \ne b$ and $j_2 \ne b$ & \multicolumn{2}{c|}{}      & than $j_1$ or$j_2$)  \\         
& where  $V \to j_1 j_2$  & \multicolumn{2}{c|}{0} &  \\
\hline
\end{tabular}
\end{center}
\end{table}

The unique kinematic signature of the VBF process is used to define event categories enriched in this production mechanism and maximize the sensitivity by reducing the SM backgrounds. Events with two small-$R$ (``tag'') jets with invariant mass $m^{\rm tag}(j,j) >$ 770\,\gev\ and pseudorapidity gap between them $|\Delta \eta^{\rm tag} (j,j)| >4.7$
are classified as VBF candidates. In case there are more than two tag-jets, the pair with the largest invariant mass is chosen. Events that fail the VBF selection are assigned to the $\mathrm{ggF/q\bar{q}}$ category. 

Events belonging to the VBF or $\mathrm{ggF/q\bar{q}}$ categories are further assigned to the merged or resolved regions as follows:
\begin{itemize}
\item {Merged signal region: the large-$R$ jet with the highest \pt\ is selected as the candidate for the hadronically decaying
$V$ boson, requiring no overlap with either of the tag-jets in the VBF category ($\Delta R(j^\mathrm{tag},J) >$ 1.0). Furthermore, the event is required to have \met > 100\,\gev\ to suppress the multijet contamination.  
 The leptonically decaying $W$ candidate is required to have a lepton--neutrino system with transverse momentum $\pt(\ell\nu) >$ 200\,\gev. A threshold of 0.2 is set on the ratio $\met / \pt(e \nu)$ in the electron channel in order to further suppress the multijet background. In the desired signal topology, the two bosons are produced from a heavy resonance decay and their transverse momenta are expected to be close to half the reconstructed resonance mass. As a result, a threshold of 0.4 (0.3) is applied to $\pt(J)/m(WV)$ and $\pt(\ell\nu)/m(WV)$ in the $\mathrm{ggF/q\bar{q}}$ (VBF) category. Furthermore, events are rejected if there is a $b$-tagged jet present with a separation of $\Delta R >$ 1.0 from the hadronically decaying $V$ candidate. The latter requirement rejects more than 70\% of background events from $t\bar{t}$ production while keeping more than 95\% of signal events, independently of the resonance mass. The remaining events are assigned to the high-purity (HP) region if the large-$R$ jet satisfies the $V$-tagging 50\%\ efficiency working point, for both the mass window and the $D_2$ variable, as defined in Section~\ref{sec:objects}. Otherwise, events are assigned to the low-purity (LP) region if the 80\%\ efficiency working point is satisfied for the large-$R$ jet. The improvement in cross-section sensitivity resulting from combining the HP and LP regions reaches up to 36\% for resonances with 5.0 \tev\ mass. The selected HP and LP events can simultaneously pass both the $WW$ and the $WZ$ selections if the large-$R$ jet passes both the $W$ and $Z$ selections.}

\item {Resolved signal region: events not satisfying the selection criteria of the merged signal region and with \met > 60\,\gev\ and $\pt(\ell\nu) >$ 75\,\gev\ are considered. The hadronically decaying $V$ candidate is formed by combining the two small-$R$ jets, excluding VBF tag-jets, with the highest \pt\ and requiring their invariant mass to be between 66 and 94 (82 and 106) \gev\ in order to be consistent with the $W$ ($Z$) boson mass. The two selected small-$R$ jets are required to have \pt > 45\,\gev\ (60\,\gev\ for the highest \pt\ jet) and the azimuthal angle separation between jets, lepton and \met\ directions must satisfy $\Delta\phi(j,\ell)>$ 1.0, $\Delta\phi(j,\met ) >$ 1.0, $\Delta\phi(j,j) <$ 1.5  and $\Delta\phi(\ell,\met) <$ 1.5. In the calculation of the $WV$ invariant mass, a $V$ mass constraint is imposed on the two small-$R$ jets by rescaling the \pt\ of the dijet system to be $\pt^{jj}\times m(V)/m(jj)$, where $\pt^{jj}$ and $m(jj)$ are the transverse momentum and the invariant mass of the dijet system respectively, and $m(V)$ is the known value of the $V$ boson mass. Studies using MC simulated events show that the mass constraint reduces the uncertainties due to the jet energy scale and results in an approximately 20\% improvement of  the resolution of the reconstructed diboson resonance mass, which ranges between 20 \gev\ and 120 \gev\ across the mass spectrum. In addition, selected events in the $\mathrm{ggF/q\bar{q}}$ (VBF) category are required to satisfy $\pt(jj)/m(WV) >$ 0.35 (0.3) and $\pt(\ell \nu)/m(WV) >$ 0.35 (0.3). Events are rejected from the $WW$ selection if both jets from the $V$ boson decay are tagged as $b$-tagged jets. Furthermore, events with one or more $b$-tagged jets, not compatible with the $V$ boson decay, are also removed. As in the merged signal region, a threshold of 0.2 is set on the ratio $\met / \pt(e \nu)$ to suppress the multijet background in the  $e \nu qq$ channel.
}

\end{itemize}

The signal efficiency times acceptance ($\epsilon \times A$), defined as the ratio of the number of signal events in the signal region to the number of generated signal events, is presented as a function of the $WV \to \ell \nu qq$ resonance mass in Figures~\ref{fig:facceptance1} and \ref{fig:facceptance0} for all the generated benchmark signals. Experimental factors, such as the detector coverage and the pile-up activity, lead to low tagging efficiency of the VBF jets resulting in small $\epsilon \times A$. Priority is given to the VBF category, using the selection outlined previously, aiming to increase the sensitivity to genuine VBF signal events that have a small signal $\epsilon \times A$ in the VBF category and the high fraction of the VBF signal that leaks in the $\mathrm{ggF/q\bar{q}}$ category. The leakage occurs due to inefficiencies related to the reconstruction and identification of the ``tag'' jets, and results in a small deterioration in sensitivity after accounting for the background. Concerning the $\epsilon \times A$ of the various analyses, the resolved analysis is more sensitive in the low mass region, while the merged analysis is more efficient in the high mass region with a relatively constant $\epsilon \times A$. In the $\mathrm{ggF/q\bar{q}}$ category, the $\epsilon \times A$ values are generally lower for the scalar signal because the two bosons are produced less centrally than for the spin-1 and spin-2 signals, and the $\pt(V)/m(WV)$ requirements reject more signal.

\begin{figure}[h]
\centering
\subfigure[][\label{fig:acceptance5}]{\includegraphics[width=0.48\linewidth]{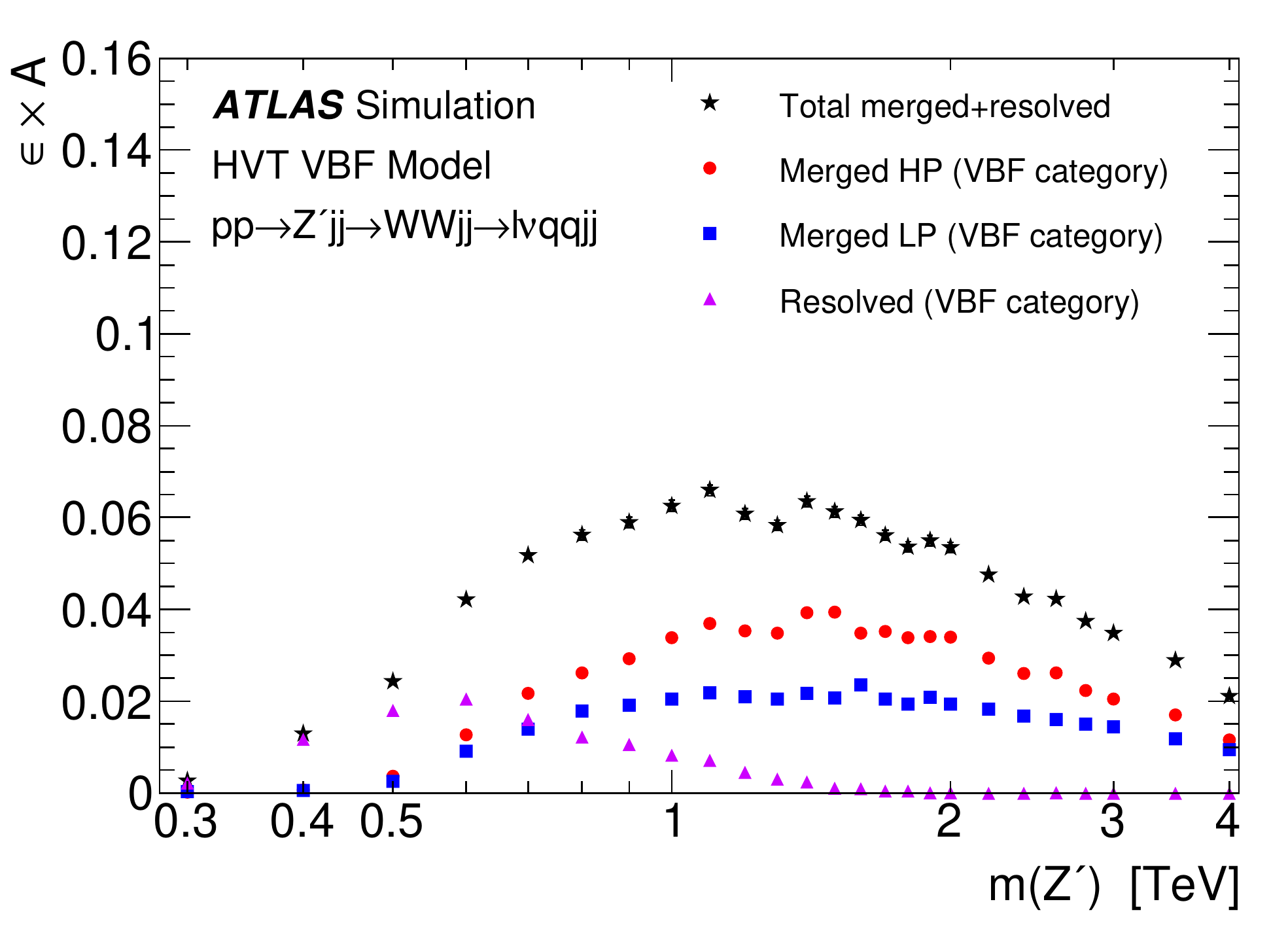}} 
\subfigure[][\label{fig:acceptance6}]{\includegraphics[width=0.48\linewidth]{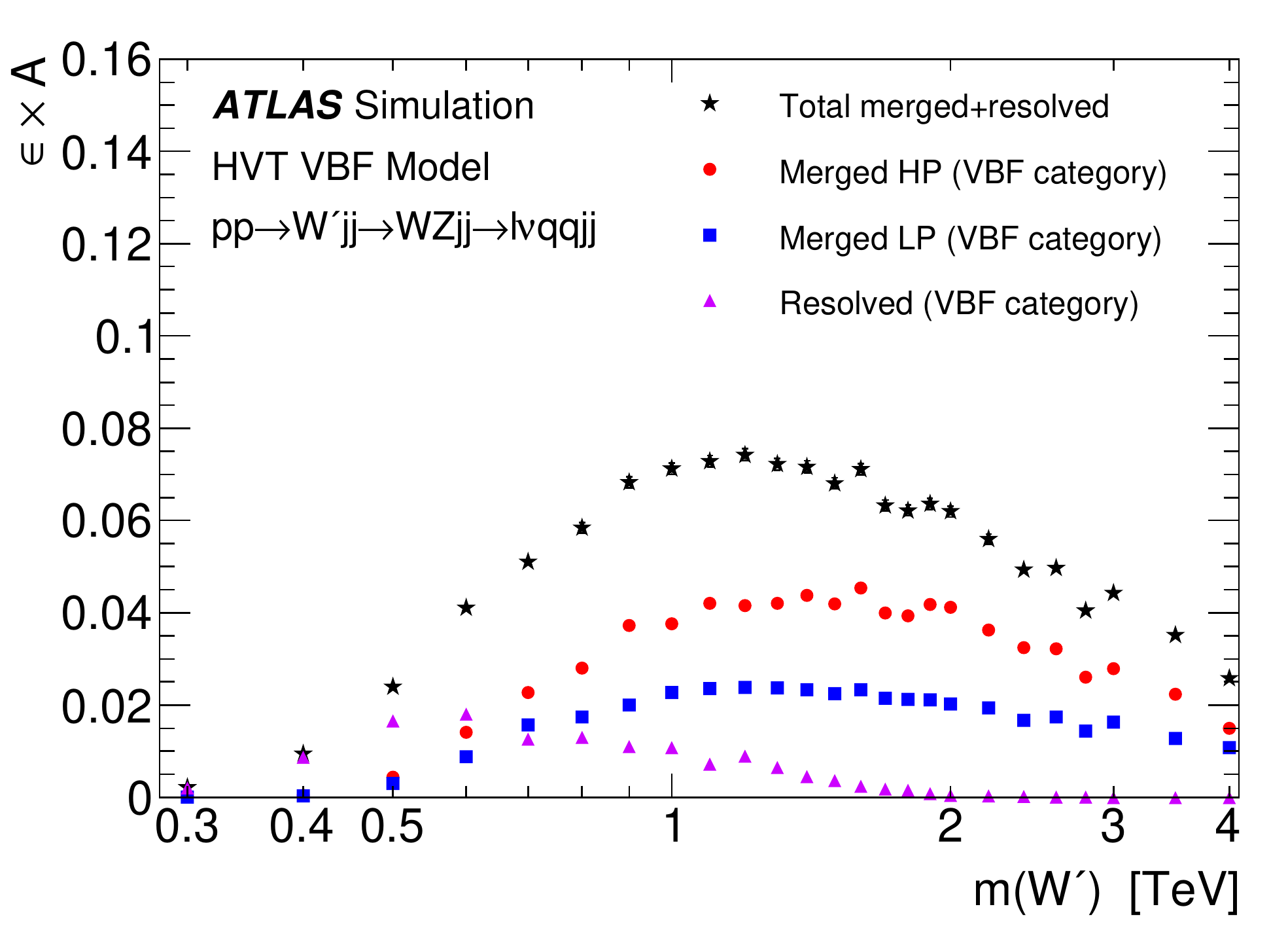}} 
\subfigure[][\label{fig:acceptance3}]{\includegraphics[width=0.48\linewidth]{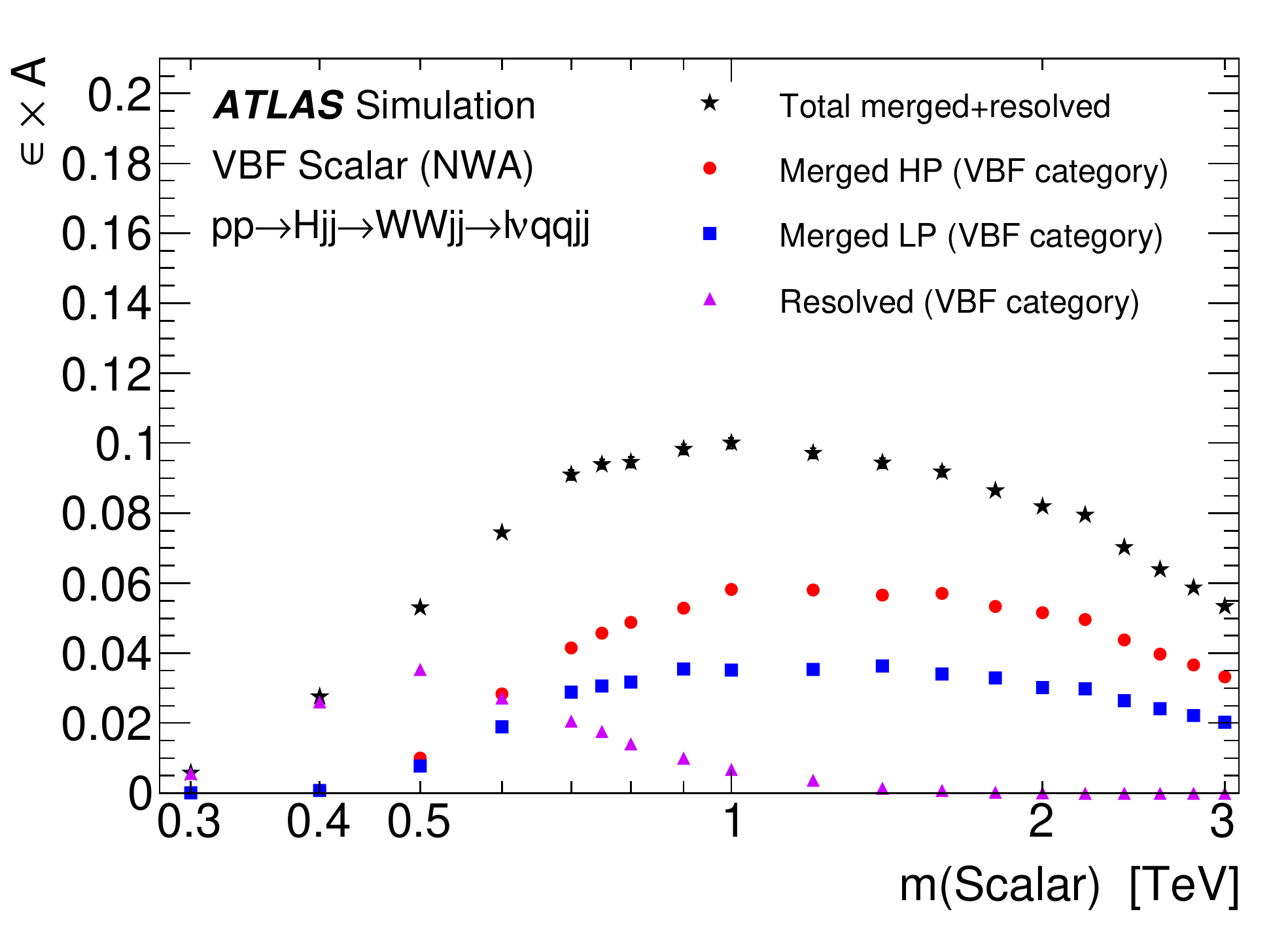}}
\caption{The product of signal efficiency ($\epsilon$) and acceptance ($A$) for signals produced via the VBF mechanism is presented in the VBF category. The $\epsilon \times A$ is shown for \subref{fig:acceptance5} HVT $WW \to \ell  \nu qq$, \subref{fig:acceptance6} HVT $WZ \to \ell  \nu qq$ and \subref{fig:acceptance3} neutral scalar signal ($H$) in the narrow-width approximation (NWA) decaying into $\ell \nu qq$ in the various analysis signal regions. It is defined as the ratio of the number of signal events reconstructed in the signal region to the number of generated signal events.}
\label{fig:facceptance1}
\end{figure}

\begin{figure}[h]
\centering
\subfigure[][\label{fig:acceptance0}]{\includegraphics[width=0.48\linewidth]{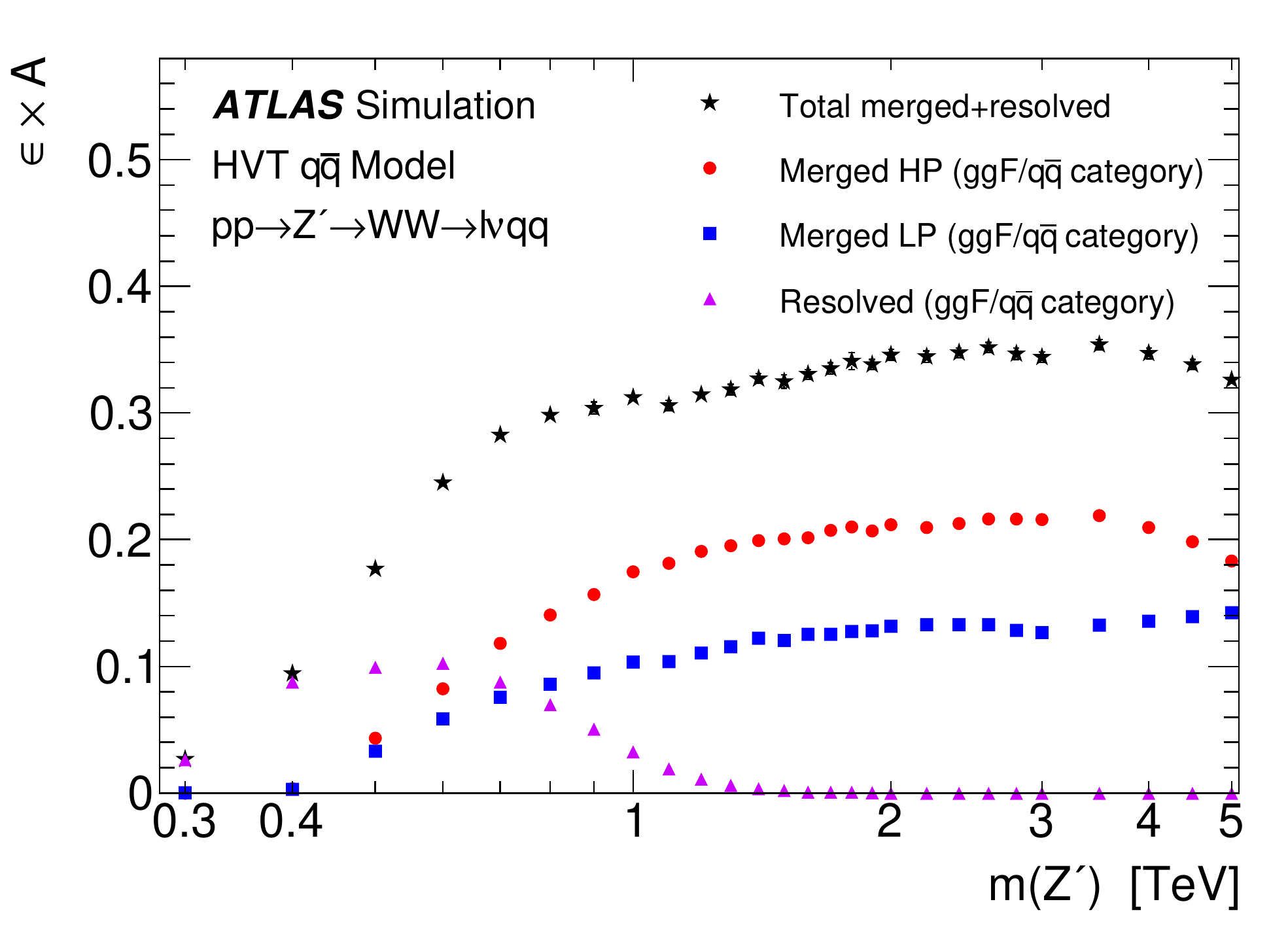}}
\subfigure[][\label{fig:acceptance1}]{\includegraphics[width=0.48\linewidth]{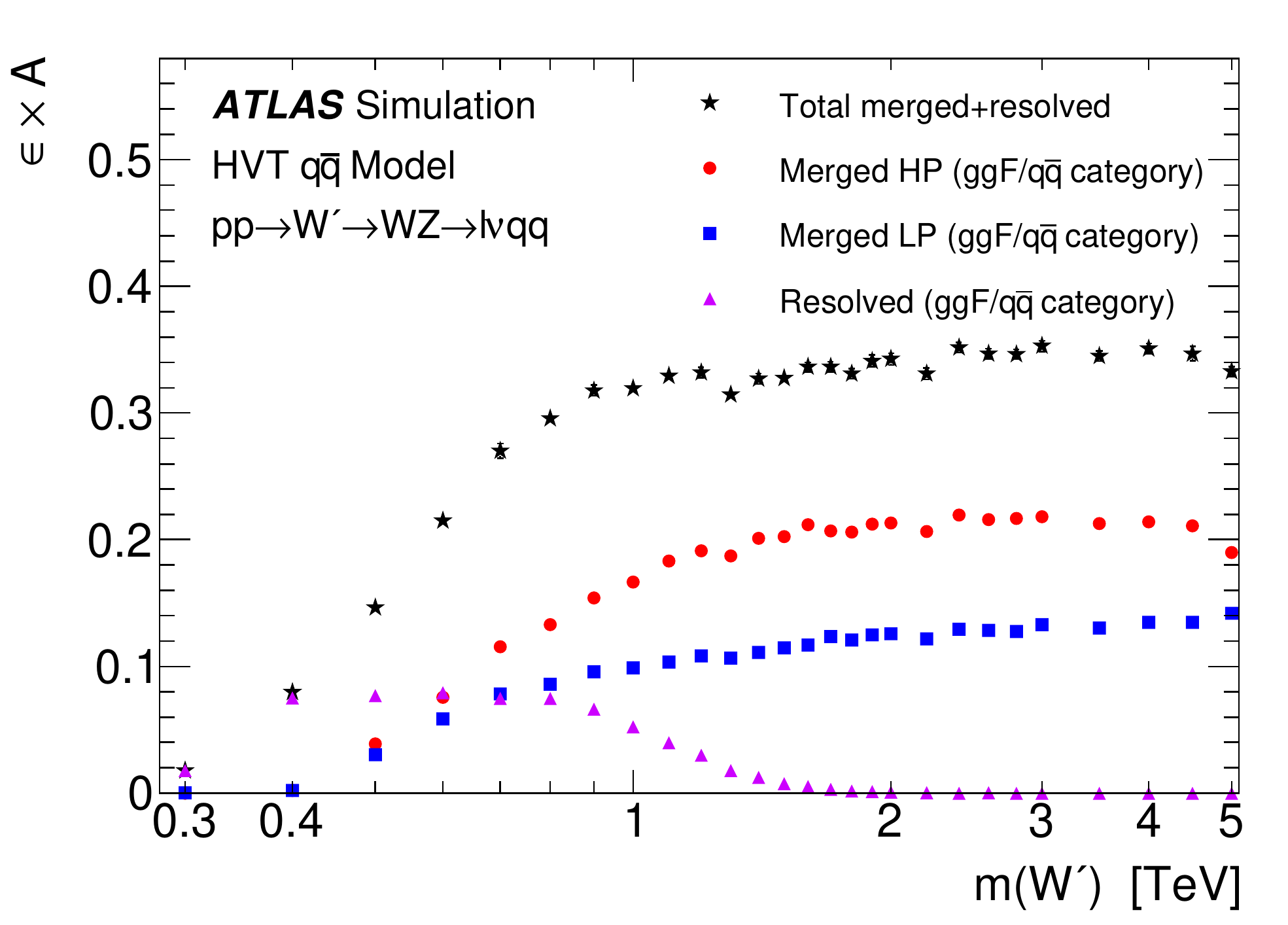}}
\subfigure[][\label{fig:acceptance2}]{\includegraphics[width=0.48\linewidth]{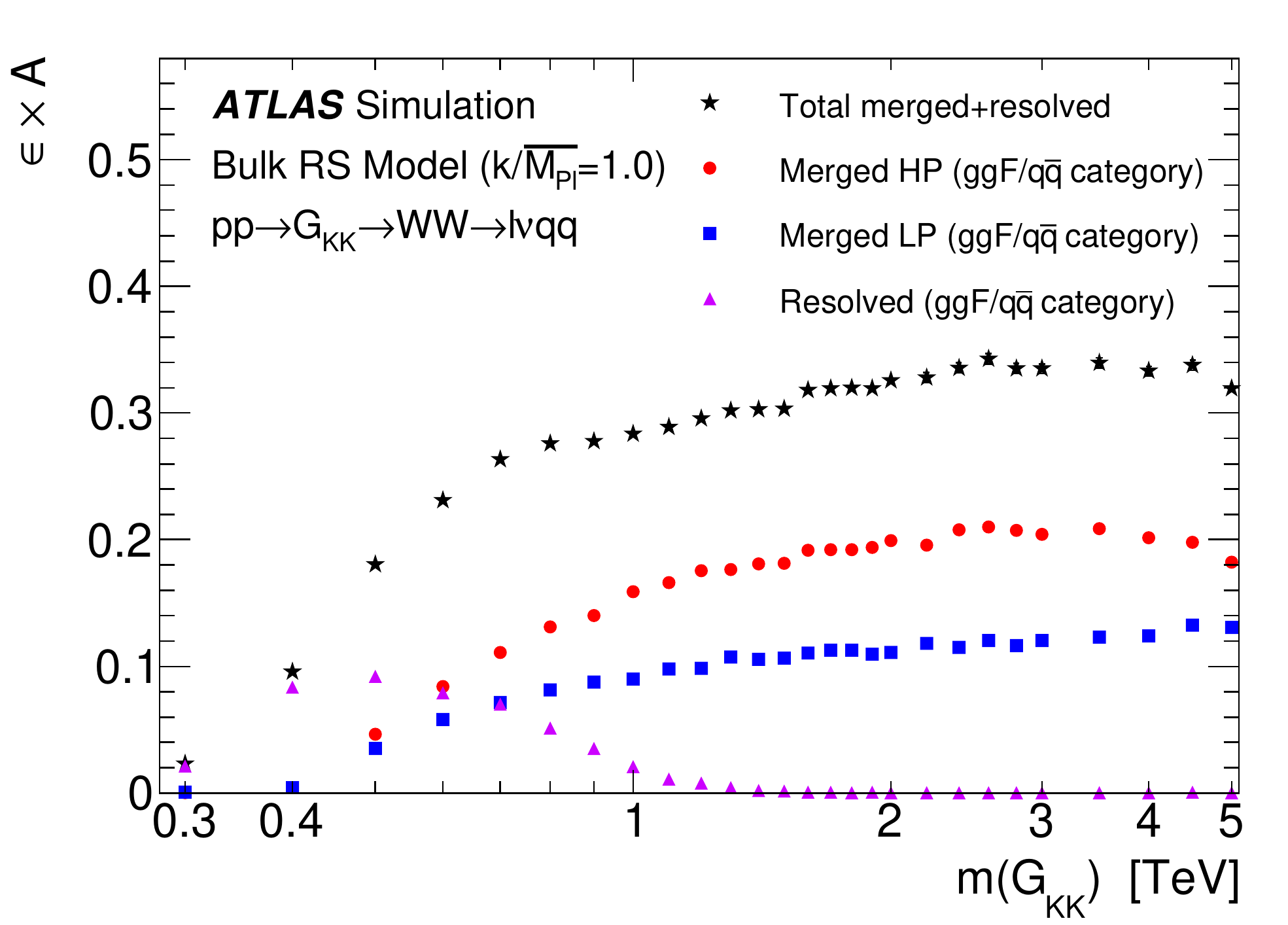}}
\subfigure[][\label{fig:acceptance4}]{\includegraphics[width=0.48\linewidth]{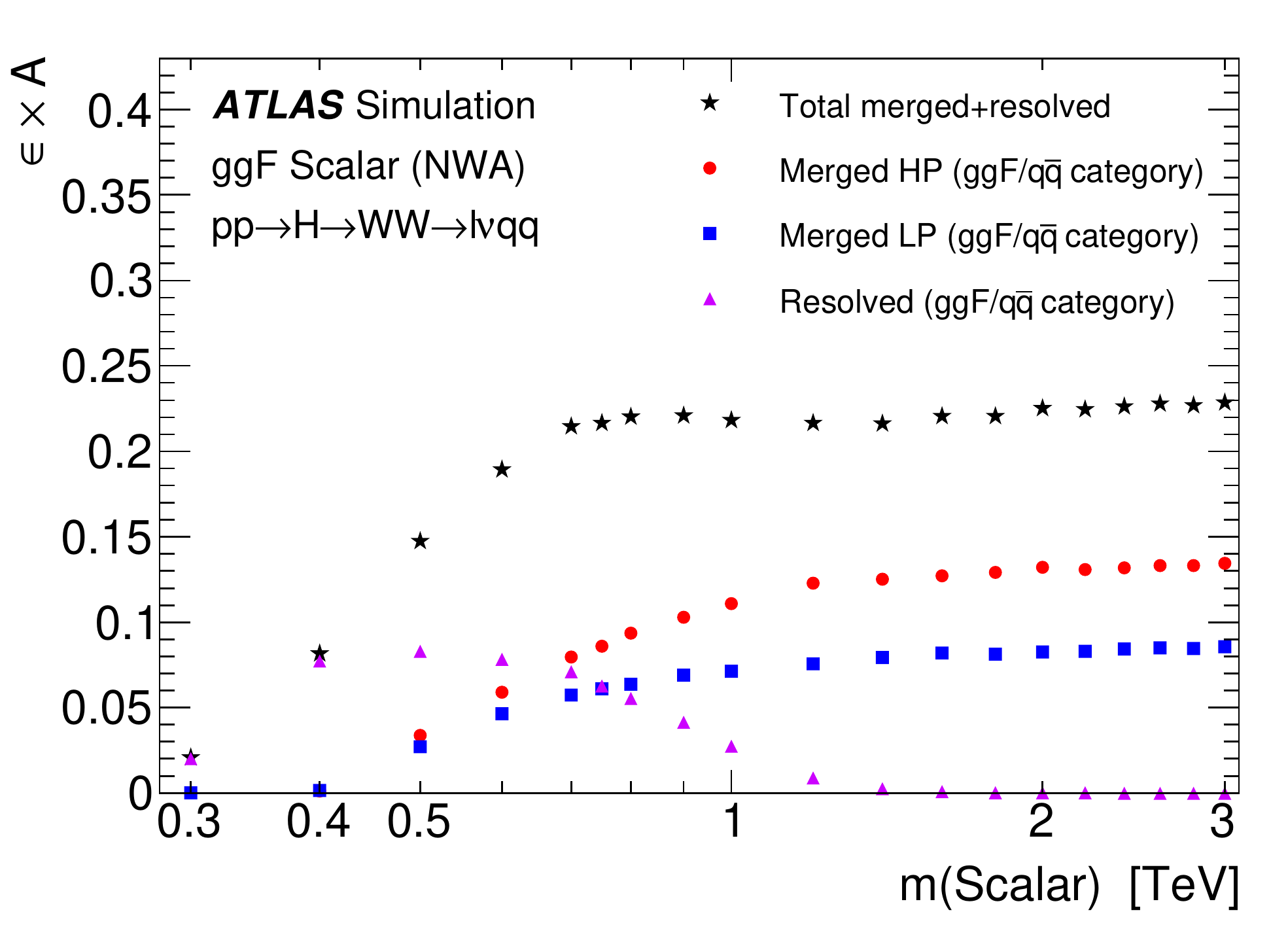}}
\caption{The product of signal efficiency ($\epsilon$) and acceptance ($A$) is presented in the $\mathrm{ggF/q\bar{q}}$ category for signals produced via ggF or $q \bar{q}$ fusion. The $\epsilon \times A$ is presented for HVT \subref{fig:acceptance0} $WW \to \ell  \nu qq$, \subref{fig:acceptance1} HVT $WZ \to \ell  \nu qq$, \subref{fig:acceptance2} RS $G_\mathrm{KK} \to \ell  \nu qq$ and \subref{fig:acceptance4} neutral scalar signal ($H$) decaying into $\ell  \nu qq$ in the narrow-width approximation (NWA) in the various analysis categories. It is defined as the ratio of the number of signal events reconstructed in the signal region to the number of generated signal events.} 
\label{fig:facceptance0}
\end{figure}

%-------------------------------------------------------------------------------
\section{Background estimation}
\label{sec:bkg}
%the ranges reflect the differences in the background composition in the $WW$ and $WZ$ signal regions from the different selections on $b$-jets, which lead to a higher $W$+jets contribution in the $WZ$ channel. 

Simulation studies indicate that the dominant background sources are $W$+jets and $t\bar{t}$ events. The $W$+jets contribution is found to be approximately 50\%, 70\% and 60\%--65\% in the high-purity, low-purity and resolved $\mathrm{ggF/q\bar{q}}$ signal regions, respectively, while the corresponding fractions in the VBF category are 40\%, 60\% and 40\%--55\%. In the resolved analysis, the $W$+jets contribution is higher in the $WZ$ channel than the $WW$ channel because of the different selections on $b$-jets. The $t\bar{t}$ contamination in the $\mathrm{ggF/q\bar{q}}$ category is estimated to be 30\% (20\%) in the high-purity (low-purity) and 25\% (30\%) in the resolved $WW$ ($WZ$) signal regions. The contribution from $t\bar{t}$ production in the VBF category is 50\%, 30\% and 35\%--50\% in the high-purity, low-purity and the resolved signal regions, respectively. Smaller background contributions arise from $Z$+jets, single-top and SM diboson production. Control regions for the high- and low-purity categories as well as the resolved category are defined for events that fail the selection criteria of the signal regions in order to estimate the dominant background contributions:
\begin{itemize}

\item {The $W$+jets control regions are formed from events satisfying the signal region selection except for the invariant mass requirement of the hadronically decaying $V$ candidate. The mass is required to be in the sideband region which is defined as $m(jj) <$ 66\,\gev\ or 106 < $m(jj)$ < 200\,\gev\ for the resolved analysis. In the merged analysis, the sideband regions are formed by events satisfying the respective $D_2$ selections but not the mass window requirement for the 80\% efficiency working point. Approximately 65\% and 77\% of the selected events are from $W$+jets production in the $\mathrm{ggF/q\bar{q}}$ category of the merged and resolved analyses, respectively. The remaining events are primarily from $t\bar{t}$ production. The contribution from $W$+jets processes is 50\% and 65\% for the merged and resolved analyses, respectively, in the VBF category.  
}
 
\item {The $t\bar{t}$ control regions are formed from events satisfying the signal region selection except for the $b$-jet requirement, which is inverted. Studies using simulated events show that 77\%--87\% of the selected $\mathrm{ggF/q\bar{q}}$ and VBF category events are from $t\bar{t}$ production and the rest are from single-top, $V$+jets or diboson production, for both the merged and the resolved event topologies. }

\end{itemize}

The shapes of the mass distributions for events from production of $W$+jets and $t\bar{t}$
are modelled using simulated events. Their normalizations are
determined from a combined fit to the events in the signal and control
regions, as detailed in Sec.~\ref{sec:results}. Concerning the subdominant background contributions from $Z$+jets, single-top and SM diboson production, simulation is used to obtain the shapes and normalizations, which are subsequently constrained within statistical, experimental and theoretical uncertainties. 

The contribution from multijet production primarily consists of events with jets or photon conversions misidentified as leptons or real but non-prompt leptons from decays of heavy-flavour hadrons. The multijet background in the merged event topology is estimated by a fit to the \met\ distribution of events that satisfy all the signal selection criteria but without any \met requirement. The shape of multijet events is obtained from an independent data control sample that satisfies the signal selection criteria except for the \met\ requirement and the lepton requirement: the leptons are required to satisfy the veto lepton selection, defined in Section~\ref{sec:objects}, but not the signal lepton selection. Contributions from other processes with prompt leptons to the control sample are subtracted from the data using samples of simulated events in the extraction
of the multijet background shape. In the fit, the normalizations of the $W$+jets and multijet components are allowed to float, with all the other
backgrounds fixed to their predicted cross sections. Following this procedure, the multijet background in the merged event topology is found to be negligible.

A fake-factor method is implemented to estimate the multijet background contribution in the resolved topology. The ``signal lepton'' control region is formed by events that have exactly one signal lepton and exactly one small-$R$ jet. The same event selection criteria are applied to the events in the ``inverted lepton'' control region except for the lepton requirement: the selected electron candidate is required to pass the ``medium'' but fail the ``tight'' requirements, and the selected muon candidate is required to fail the nominal but pass a looser isolation requirement. The fake-factor is defined as the ratio of the number of events in the signal lepton control region to the number of events in the inverted lepton control region, after subtracting contributions from prompt leptons as estimated by the simulation. The fake-factor is calculated as a function of the lepton \pt\ and $\eta$, and \met. It is subsequently used to reweight a sample of events selected with the inverted lepton selection, as previously described, that satisfy the rest of the signal region selection.

%-------------------------------------------------------------------------------
\section{Systematic uncertainties}
\label{sec:systematics}
Systematic uncertainty sources impacting the search can be divided into four categories: experimental uncertainties related to the detector or reconstruction algorithms, uncertainties in the estimations of background contributions, uncertainties in modelling the signal and statistical uncertainties in the MC predictions. Two kinds of background uncertainties are provided, normalization and shape uncertainties. Normalization uncertainties are extracted from data and MC simulation comparisons, while shape uncertainties are accounted for by varying MC parameters.

Modelling uncertainties affecting the shape of the final mass discriminant are estimated for the $W$+jets background. These include uncertainties in the renormalization and factorization scales, the CKKW~\cite{Hoeche:2009rj,Catani:2001cc} matching scales, the resummation scale, the PDF and $\alpha_\mathrm{s}$. The scale uncertainties are obtained by doubling and halving the corresponding parameters in the nominal generator. Potential systematic uncertainties due to choices of parton shower and matrix element implementations are estimated by comparing the nominal MC samples to the alternative samples generated using \textsc{MadGraph}.

The uncertainty in the shape of the $m(WV)$ distribution from the $t\bar{t}$ background is estimated by comparing the nominal \textsc{Powheg+Pythia} sample to the alternative samples described in Section~\ref{sec:simul}. The factorization and renormalization scales of the nominal generator are varied, in a similar manner as the $W$+jets parameters, and their difference from the nominal sample is also applied as a systematic uncertainty. 

The SM diboson production cross section is fixed to the inclusive next-to-leading-order calculation with a 30\% systematic uncertainty in the normalization. The $m(WV)$ distribution shape uncertainty of the diboson background is estimated by comparing the predictions based on the alternative \textsc{Powheg-Box} MC samples to those of the nominal \textsc{Sherpa} MC samples.

Systematic uncertainties in the multijet background estimate are only considered in the resolved analysis, as this background contribution in the merged analysis is negligible. These are obtained by varying the lepton or isolation selection used in the fake-factor calculation. In addition, the statistical uncertainties of the measured fake-factors and the systematic uncertainties in the prompt lepton contribution in the measurement of the fake-factors, are taken into account in the estimation of systematic uncertainties of the multijet background modelling. The effect of this uncertainty is found to be marginal in the fit.

Experimental uncertainties related to leptons, jets and \met are considered, affecting the shape and normalization of both the background and the signal distributions. These are estimated for the trigger efficiencies, the energy scale and resolution of small-$R$ jets~\cite{PERF-2016-04} and large-$R$ jets~\cite{ATL-PHYS-PUB-2015-033}, lepton identification, reconstruction and isolation efficiencies, lepton momentum scales and resolutions~\cite{ATLAS-CONF-2016-024,ATL-PHYS-PUB-2016-015,PERF-2015-10}, $b$-tagging efficiency and misidentification rates~\cite{PERF-2012-04,ATL-PHYS-PUB-2016-012}, 
and missing transverse momentum resolution~\cite{ATL-PHYS-PUB-2015-023}. 

For central small-$R$ jets (|$\eta$| < 2.0), the total relative uncertainty in the jet energy scale~\cite{PERF-2016-04} ranges from about 6\% for jets with \pt\ of 25 \gev\ to about 2\% for \pt\  of 1000 \gev. The  uncertainty in the small-$R$ jet energy resolution ranges from 10\%--20\% for jets with \pt of 20 \gev\ to less than 5\% for jets with \pt > 200 \gev. 

The uncertainties in the scale of the $D_2$ variable and in the large-$R$ jet energy and mass are estimated by
comparing the ratio of calorimeter-based to track-based energy and mass measurements
in dijet data and simulation~\cite{ATL-PHYS-PUB-2015-033}. These uncertainties range between 2\% and 5\%. An uncertainty of 2\% is assigned to the large-$R$ jet energy resolution and uncertainties of 20\% and 15\% are assigned to the resolution of the large-$R$ jet mass and $D_2$, respectively.

The dominant uncertainties in the signal acceptance arise from the choice of PDF and the uncertainty in the amount of initial- and final-state radiation (ISR and FSR, respectively) in simulated events. The cross section obtained with the nominal PDF set is compared to those of the MMHT 2014 PDF~\cite{Harland-Lang:2014zoa} and CT14 PDF~\cite{PhysRevD.93.033006} to derive the uncertainties in the acceptance. The prescription in Ref.~\cite{Butterworth:2015oua} is followed and the envelope of the uncertainties associated to the three PDF sets is used. The ISR/FSR contributions are computed by varying the parton shower and multi-parton interaction parameters following the prescription in Ref.~\cite{ATL-PHYS-PUB-2014-021}.

The uncertainty in the combined 2015+2016 integrated luminosity is 3.2\%. It is derived, following a methodology similar to that detailed in Ref.~\cite{Aaboud:2016hhf}, from a preliminary calibration of the luminosity scale using $x$--$y$ beam-separation scans performed in August 2015 and May 2016. This uncertainty is applied to the yields predicted by the simulation.

%-------------------------------------------------------------------------------
\section{Results}
\label{sec:results}
\label{sec:results}
The results are extracted by performing a simultaneous binned maximum-likelihood fit to the $m(WV)$ distributions in the signal regions and the $W$+jets and $t \bar{t}$ control regions. The $WW$ and $WZ$ channels are treated individually, without combining their respective regions. A test statistic based on the profile likelihood ratio~\cite{Cowan:2010js} is used
to test hypothesized values of the global signal-strength factor ($\mu$), separately for each model considered. The likelihood is defined as the product of the Poisson likelihoods for all signal and control regions for a given production mechanism category and channel ($WW$ or $WZ$), simultaneously for the electron and muon channels. The fit includes six contributions, corresponding to $W$+jets, $t \bar{t}$, single-top, $Z$+jets, diboson and multijet events. The main background sources, namely $W$+jets and $t \bar{t}$, are constrained by the corresponding control regions and are treated as uncorrelated among the resolved and merged signal regions. For each of these backgrounds, a normalization factor, defined as the ratio of the number of simulated events after the fit to the number of simulated events before the fit, is derived and the results are collectively presented in Table~\ref{tab:norm_factors}. In all regions and categories, the normalization factors are found to be compatible with 1.0. 

Systematic uncertainties are taken into account as constrained nuisance parameters with Gaussian or log-normal distributions. For each source of systematic uncertainty, the correlations across bins of $m(WV)$ distributions and between different kinematic regions, as well as those between signal and background, are taken into account. The number of bins and the bin widths in each signal region are optimized according to the expected background event distribution and detector resolution. In the merged region, the diboson invariant mass range extends from 500\,\gev\ to 5000\,\gev\ divided into twenty (eleven) bins in the $\mathrm{ggF/q\bar{q}}$ (VBF) category. The resolved region is covered by ten (nine) bins of varying width in the $\mathrm{ggF/q\bar{q}}$ (VBF) category, beginning at 300\,\gev\ and ending at 1500\,\gev, due to the selection efficiency. In all regions, the overflow events are included in the last bin. The $m(WV)$ distributions are presented in Figures~\ref{fig:postFit:VBF} and \ref{fig:postFit:ggF} after the VBF and $\mathrm{ggF/q\bar{q}}$ categorizations, respectively, for the merged and the resolved regions.   

The list of leading sources of uncertainty in the best-fit $\mu$ value is given in Table~\ref{tab:NPranks} together with their relative importance ($\Delta \mu / \mu$). The values are quoted separately for the VBF and $\mathrm{ggF/q\bar{q}}$ categories, and for the case of high and low mass signal samples, for which the merged and resolved topologies reach the highest sensitivity respectively. The largest systematic uncertainties are related to the background modelling and jet measurements and these are most important at lower masses.
%Examples from the high and low mass regions, where the merged and the resolved topologies reach the highest sensitivity respectively, as well as the $\mathrm{ggF/q\bar{q}}$ and VBF categories, are included. 

Exclusion limits are calculated using the ${\mathrm{CL}}_\mathrm{s}$ method~\cite{Read:2002hq}, in the asymptotic approximation, at the 95\% confidence level (CL) for resonance masses below 1.0 (1.6) \tev\ in the VBF ($\mathrm{ggF/q\bar{q}}$) category. For higher masses, the small number of expected events makes the asymptotic approximation imprecise and the limits are calculated using pseudo-experiments. The limits are calculated by fitting the merged high- and low-purity signal regions simultaneously with the corresponding resolved region. The calculation is performed separately in each final state, $WW$ or $WZ$, and the largest local excess observed is approximately 2.7~$\sigma$, which is not significant. The observed and expected upper limits on the cross sections for all generated benchmark signal models are shown in Figures~\ref{fig:limitVBF} and~\ref{fig:limitggF} for the VBF and $\mathrm{ggF/q\bar{q}}$ categories respectively. Because of the small deterioration in sensitivity after accounting for the background and the unknown ratio of the various production mechanisms in the models that are considered, the interpretation in the VBF ($\mathrm{ggF/q\bar{q}}$) category assumes there is no signal leakage from $\mathrm{ggF/q\bar{q}}$ (VBF) processes. Table~\ref{tab:limres} summarizes exclusion limits on the mass for the various signal hypotheses as extracted from the $\mathrm{ggF/q\bar{q}}$ category. For signal produced via the VBF mechanism and all scalar signals, only upper limits on the cross sections are set.

\begin{table}[]
\centering
\caption{Normalization factors, defined as the ratio of the number of fitted events to the number of predicted events from simulation, of the main background sources, namely $W$+jets and $t \bar{t}$, in the VBF and $\mathrm{ggF/q\bar{q}}$ categories. The quoted uncertainties incorporate statistical and systematic uncertainties. \label{tab:norm_factors}} 
\vspace{5pt}
\begin{tabular}{|c|c|cc|cc|}
\hline
Category & Signal Region & \multicolumn{2}{c|}{$WW$ Selection}&  \multicolumn{2}{c|}{$WZ$ Selection}\\
& &  $W$+jets & $t \bar{t}$ &  $W$+jets & $t \bar{t}$\\
\hline
\multirow{2}{*}{VBF} & Merged   & 0.89 $\pm$ 0.18  & 1.21 $\pm$ 0.18      & 0.84 $\pm$ 0.16  & 1.10 $\pm$ 0.17  \\
& Resolved & 1.13 $\pm$ 0.25  & 1.22 $\pm$ 0.18                          & 1.08 $\pm$ 0.25  & 1.21 $\pm$ 0.17 \\
\hline
\multirow{2}{*}{$\mathrm{ggF/q\bar{q}}$} & Merged   & 0.95 $\pm$ 0.06  & 1.03 $\pm$ 0.06      & 0.97 $\pm$ 0.06  & 1.00 $\pm$ 0.06  \\
& Resolved                     & 1.06 $\pm$ 0.08  & 1.02 $\pm$ 0.05      & 1.06 $\pm$ 0.08  & 1.00 $\pm$ 0.05 \\
\hline
\end{tabular}
\end{table}

\begin{table}[]
\centering
\caption{
Dominant relative uncertainties in the signal-strength parameter ($\mu$) of hypothesized HVT signal production with $m(Z') =$ 1200 \gev\ and $m(W') = $ 500 \gev\ in the VBF category, and $m(W') = $  2000 \gev\ and $m(Z') =$ 500 \gev\ in the $\mathrm{ggF/q\bar{q}}$ category, assuming that the production cross sections equal the expected 95\% CL upper limits of 0.012~pb, 0.7~pb, 0.005~pb and 0.5~pb, respectively. The impact of the several other sources of systematic uncertainty remains significant, however they are not included in the table as subdominant with respect to those quoted. The effect of the statistical uncertainty on the signal and background  samples is also shown. The large-$R$ jet kinematic uncertainties arise from jet-related reconstruction uncertainties that can be dominant in the low $m(WV)$ region because of the merged analysis priority in the event categorization. The scale uncertainty of the $t\bar{t}$ background includes the uncertainties of the factorization and renormalization scales of the nominal generator. The scale uncertainty of the $W$+jets background includes the uncertainties in the renormalization and factorization scales, the CKKW matching scales, and the resummation scale. The cross section uncertainties for the $W$+jets and $t\bar{t}$ backgrounds are constrained by the statistical uncertainty of the corresponding control data.
} 

\vspace{7pt}
\label{tab:NPranks}
\begin{tabular}{lr|lr}
\hline
\hline
\multicolumn{4}{c}{VBF Category} \\
\hline
 \multicolumn{2}{c}{$m(Z')=1200\ \gev$} &         \multicolumn{2}{c}{$m(W')=500\ \gev $}  \\\hline
Source & $\Delta \mu / \mu$ [\%] & Source & $\Delta \mu / \mu$ [\%] \\
\hline
 MC statistical uncertainty          &  15& MC statistical uncertainty              & 16 \\  
 Large-$R$ jets mass resolution &  5 & $W$+jets: cross section              & 10 \\ 
 $W$+jets: PDF choice            &  5 & Multijet \met\ modelling             & 10 \\
 $t\bar{t}$: alternative generator      &  5 & Small-$R$ jets energy resolution    &  9 \\
 $W$+jets: cross section         &  5 & SM diboson cross section            &  8 \\ 
 $t\bar{t}$: scales                   &  4 & $t\bar{t}$: cross section                   &  7 \\
\hline
Total systematic uncertainty  &  24 & Total systematic uncertainty     &  40 \\
Statistical uncertainty          &  52 & Statistical uncertainty             &  30 \\
\hline
 \multicolumn{4}{c}{}\\
 \multicolumn{4}{c}{}\\
\hline
 \multicolumn{4}{c}{$\mathrm{ggF/q\bar{q}}$ Category} \\
\hline
 \multicolumn{2}{c}{$m(W')=2000\ \gev$} &         \multicolumn{2}{c}{$m(Z')=500\ \gev $}  \\\hline
Source & $\Delta \mu / \mu$ [\%] & Source & $\Delta \mu / \mu$ [\%] \\
\hline
MC statistical uncertainty              &  12& Large-$R$ jets kinematics           & 17 \\
$W$+jets: generator choice           &  8 & MC statistical uncertainty              & 12 \\
$W$+jets: scale                   &  5 & $t\bar{t}$: scale                        & 11 \\
SM diboson normalization            &  4 & SM diboson cross section            & 10 \\
Large-$R$ jets mass resolution      &  4 & $W$+jets: alternative generator      & 10 \\
Large-$R$ jets $D_2$ resolution   &  4 & $W$+jets: scale                   &  9 \\
\hline 
Total systematic uncertainty      & 20  & Total systematic uncertainty     & 42 \\
Statistical uncertainty              & 50  & Statistical uncertainty             & 18 \\
\hline \hline
\end{tabular}
\end{table}

\begin{figure}[]
\centering
%HP
\vspace{-0.82cm}
\subfigure[][\label{fig:bpm0a}]{ \includegraphics[width=0.4\hsize]{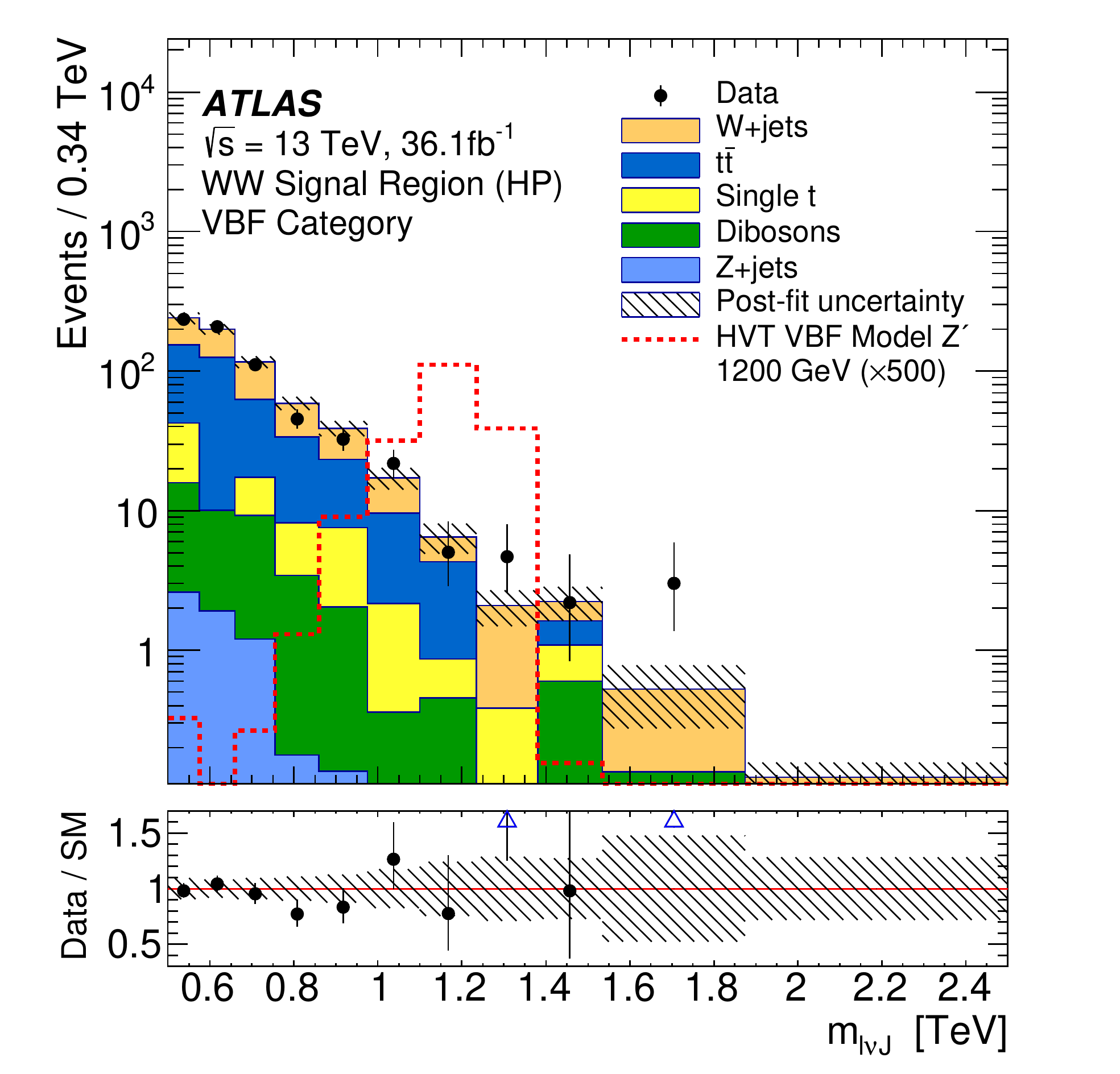}}
\subfigure[][\label{fig:bpm0b}]{ \includegraphics[width=0.4\hsize]{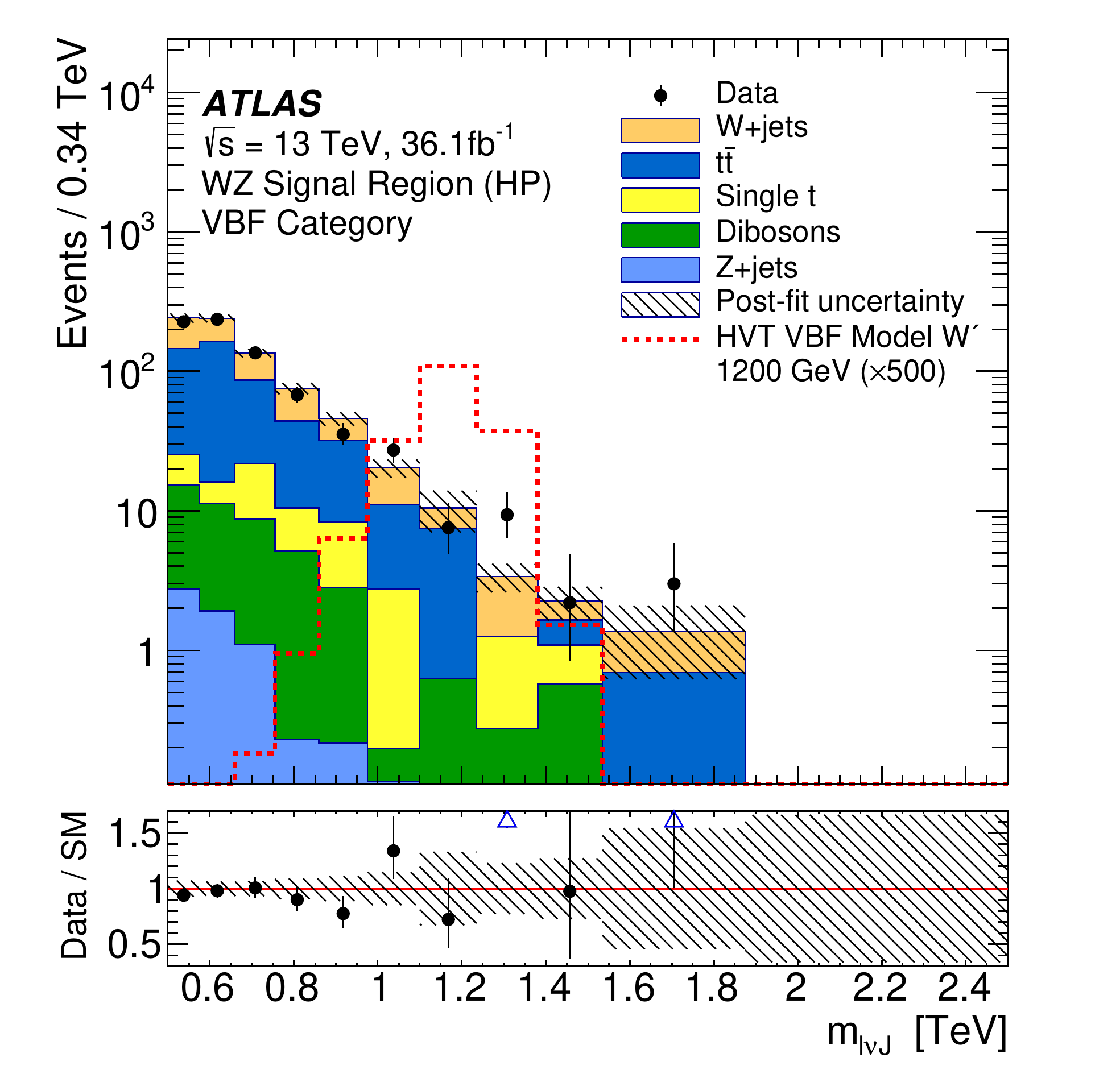}} 
%LP
\vspace{-0.34cm}
\subfigure[][\label{fig:bpm1a}]{ \includegraphics[width=0.4\hsize]{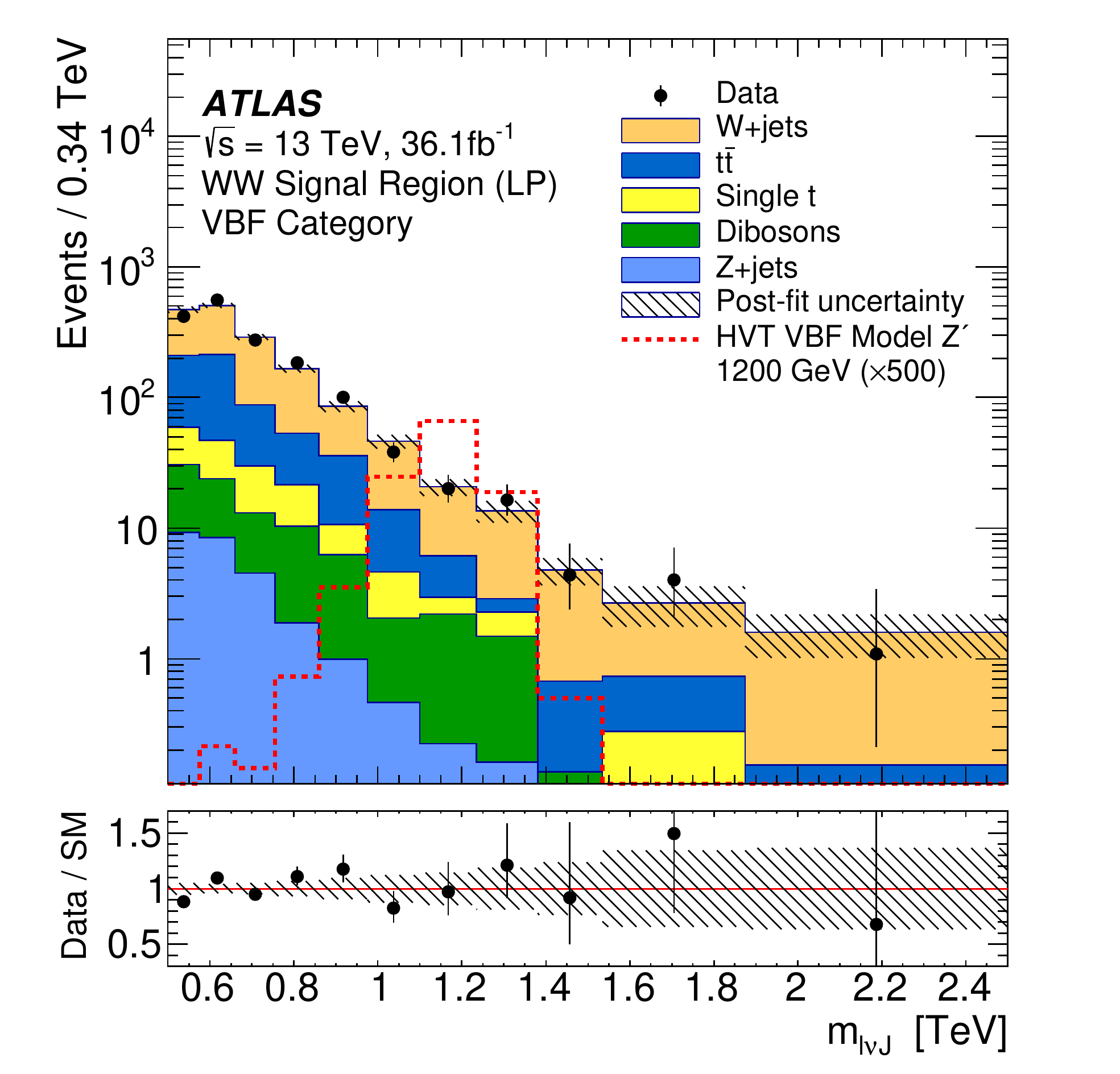}}
\subfigure[][\label{fig:bpm1b}]{ \includegraphics[width=0.4\hsize]{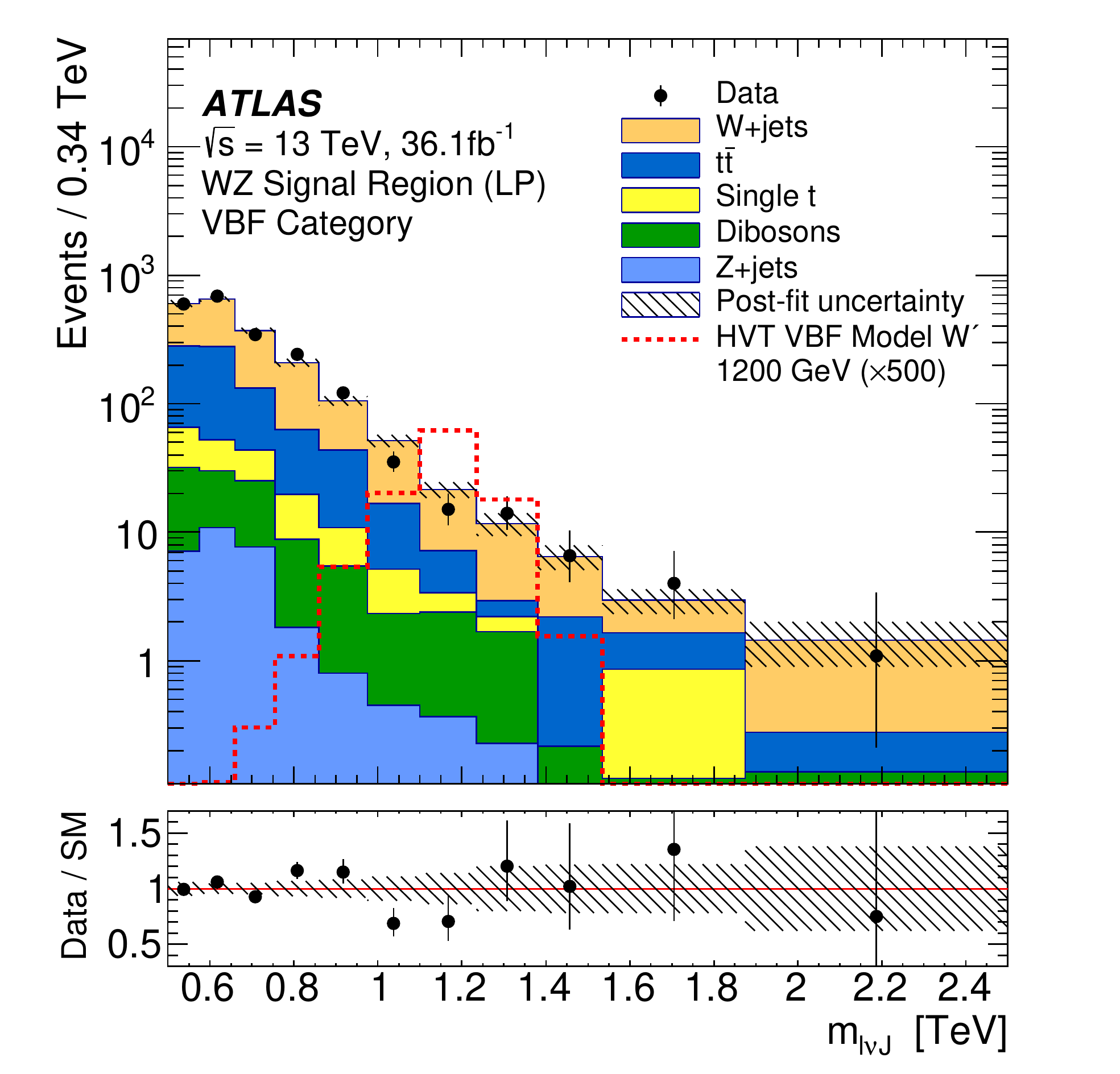}} 
\vspace{-0.4cm}
\subfigure[][\label{fig:pm2a}]{ \includegraphics[width=0.4\hsize]{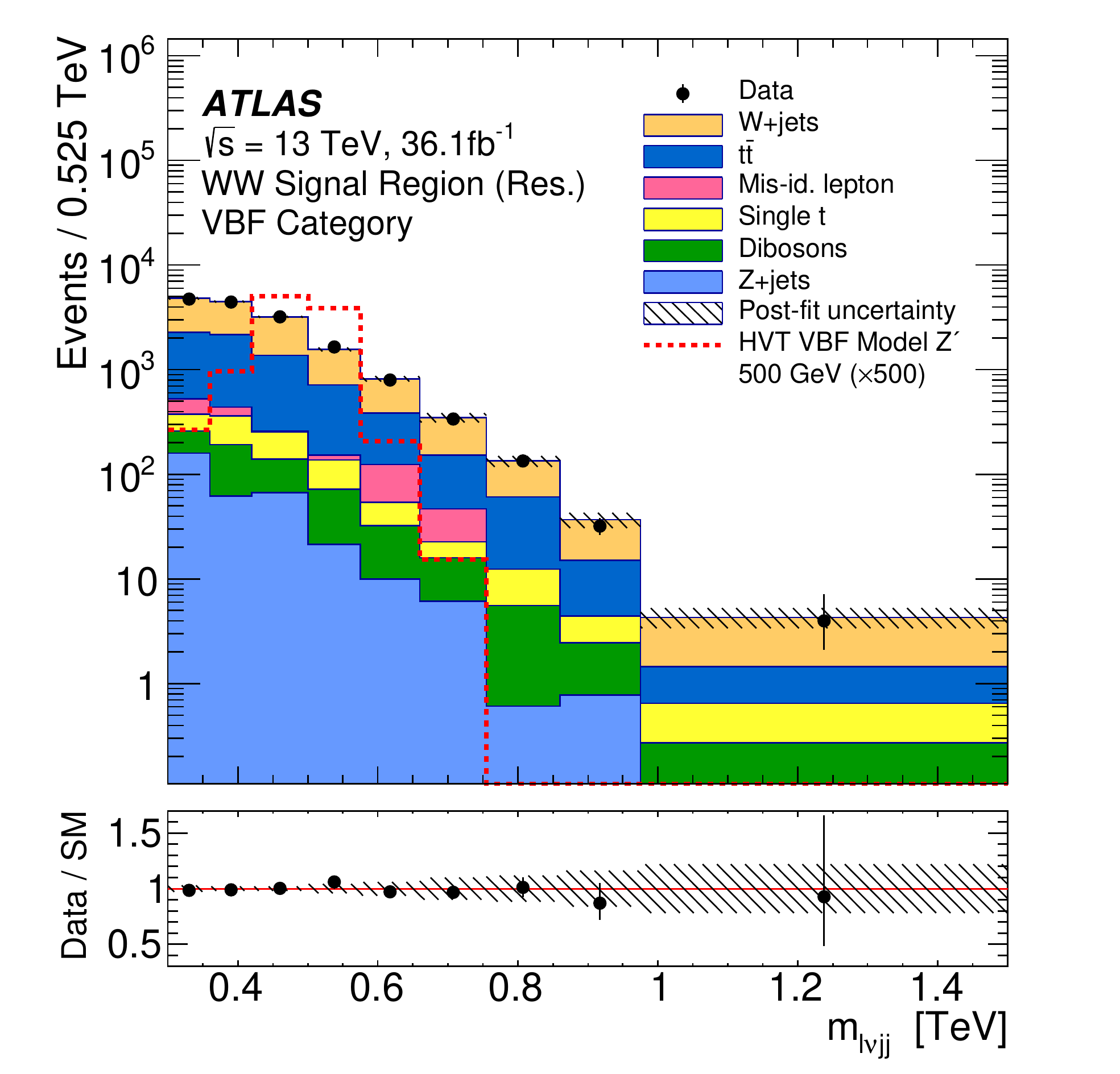}}
\subfigure[][\label{fig:pm2b}]{ \includegraphics[width=0.4\hsize]{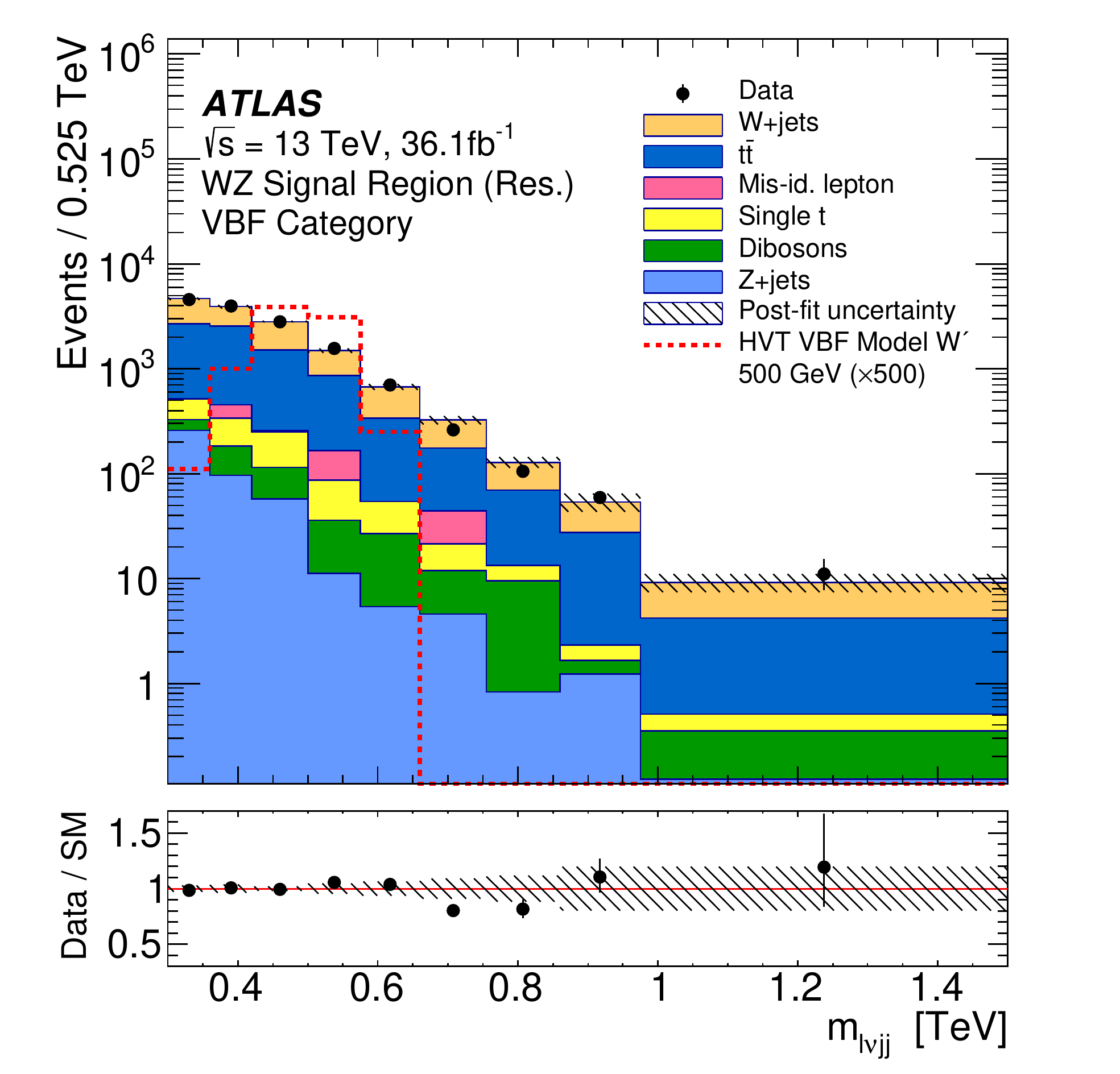}}
\caption{Post-fit signal region $m(WV)$ distributions in the VBF category. The merged high-purity (HP) sample of \subref{fig:pm0a} $WW$ and \subref{fig:pm0b} $WZ$ events, the merged low-purity (LP) sample of \subref{fig:pm1a} $WW$ and \subref{fig:pm1b} $WZ$ events and the resolved (Res.) sample of \subref{fig:pm2a} $WW$ and \subref{fig:pm2b} $WZ$ events are presented. The expected background is shown after the profile likelihood fit to the data, and signal predictions are overlaid, normalized to the cross sections indicated in the legends. The VBF HVT signal at 1200 \gev\ is presented for the merged analysis, while the 500 \gev\ signal is shown in the resolved topology. The band denotes the statistical and systematic uncertainty in the background after the fit to the data. The lower panels show the ratio of the observed data to the estimated SM background. The distribution of events is shown per mass interval corresponding to the penultimate bin width, while the overflow events are included in the last bin.}
\label{fig:postFit:VBF}
\end{figure}

\begin{figure}[]
\centering
\vspace{-0.82cm}
%HP
\subfigure[][\label{fig:pm0a}]{ \includegraphics[width=0.4\hsize]{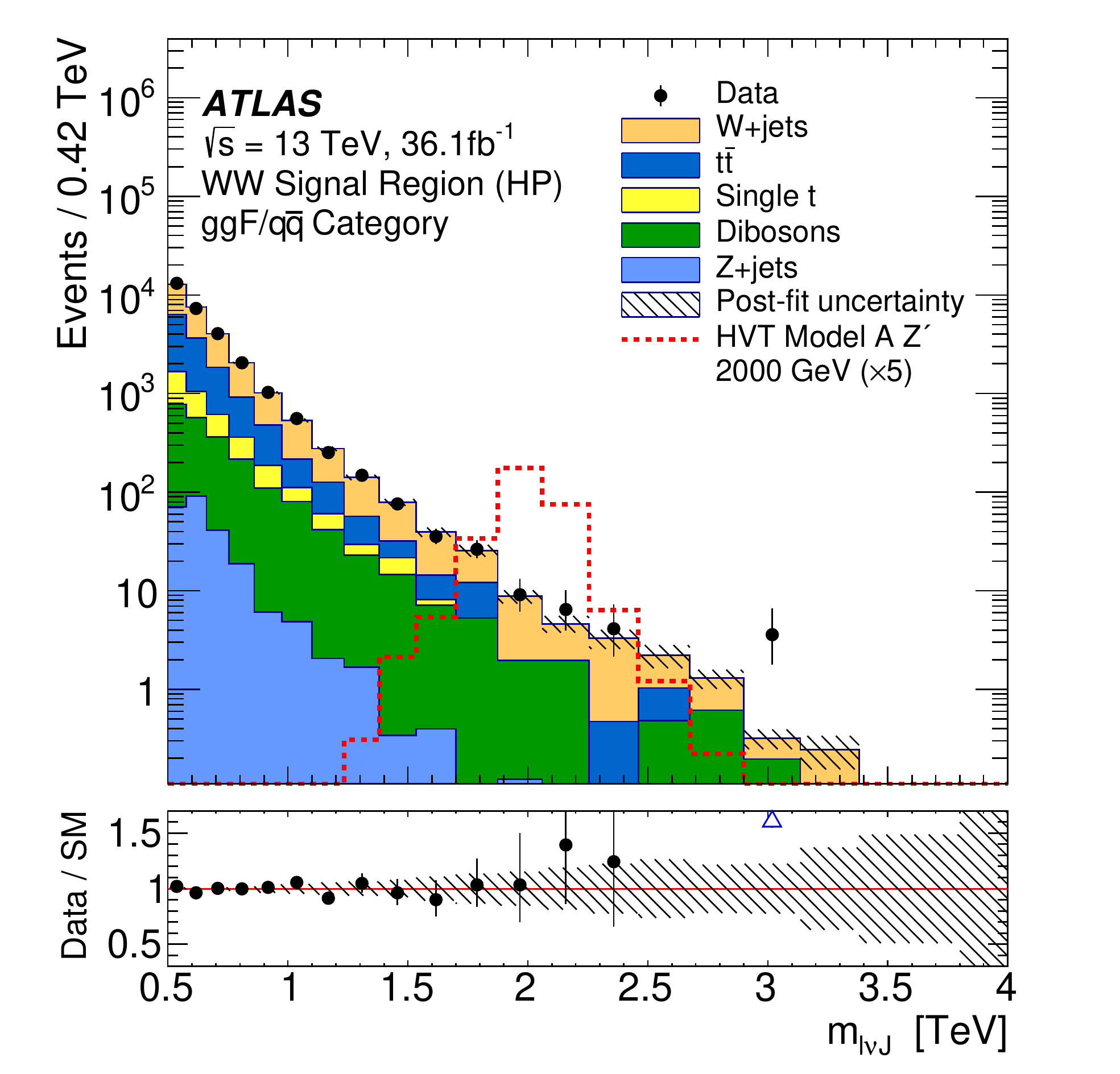}}
\subfigure[][\label{fig:pm0b}]{ \includegraphics[width=0.4\hsize]{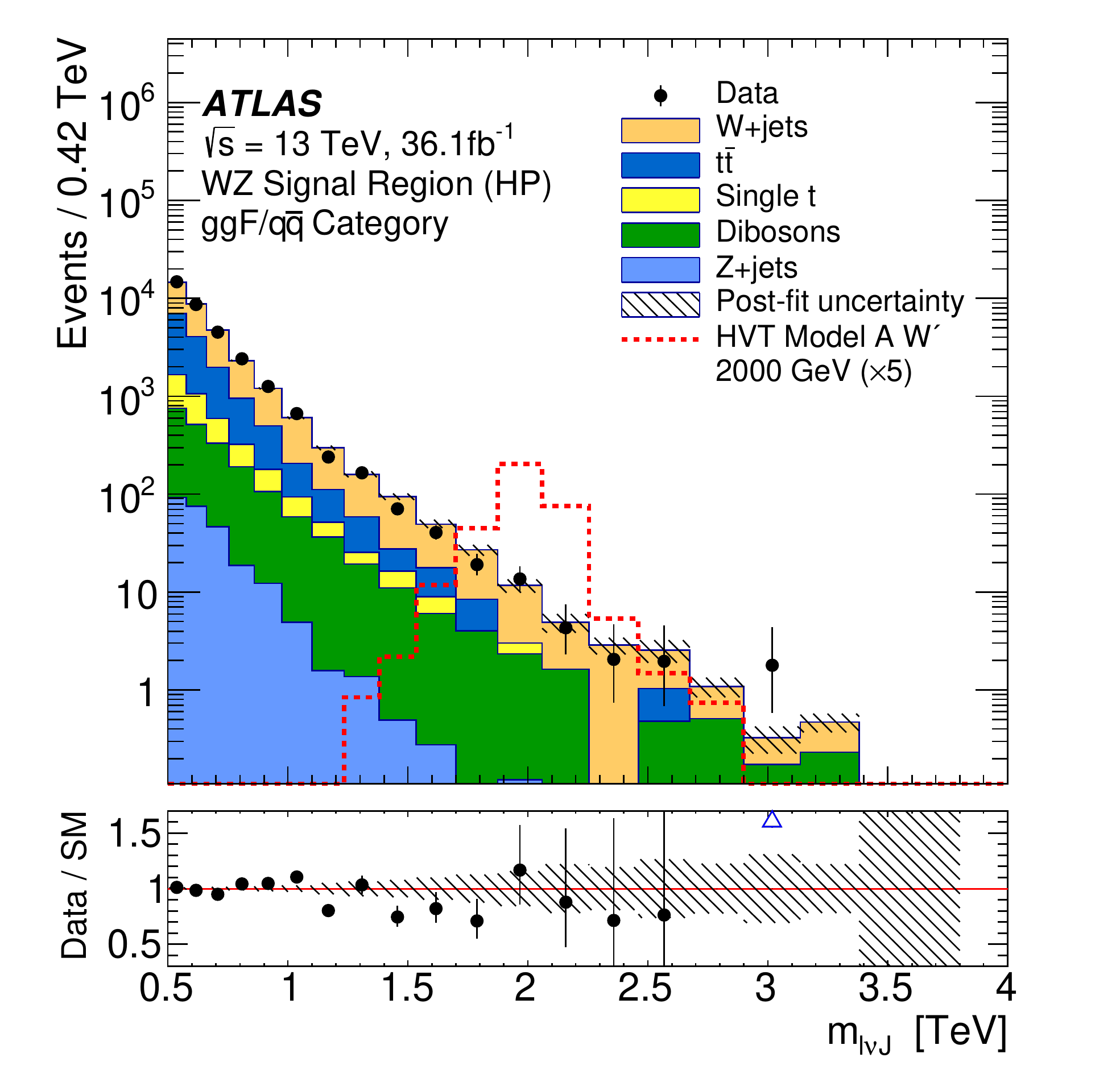}} 
%LP
\vspace{-0.34cm}
\subfigure[][\label{fig:pm1a}]{ \includegraphics[width=0.4\hsize]{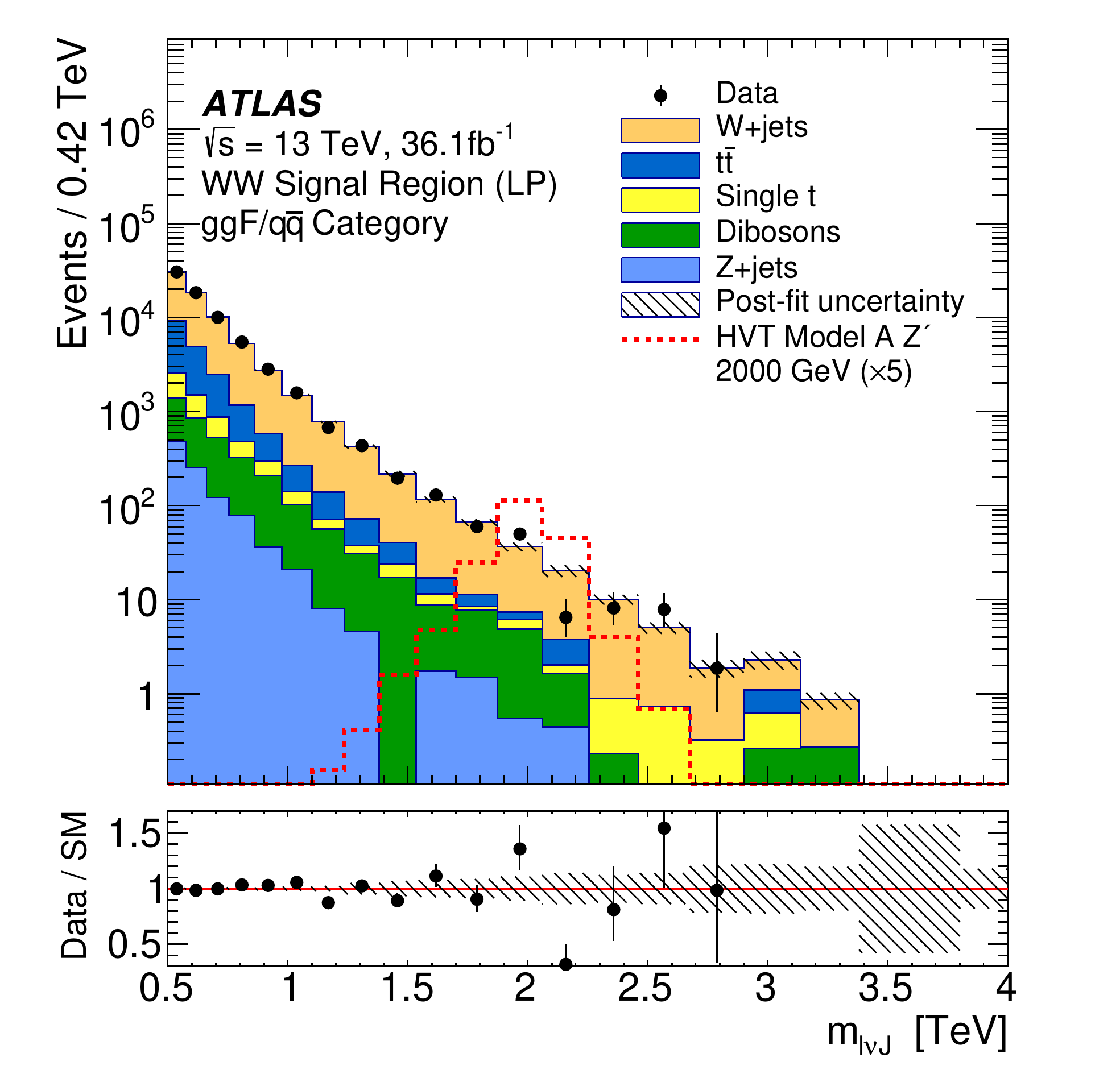}}
\subfigure[][\label{fig:pm1b}]{ \includegraphics[width=0.4\hsize]{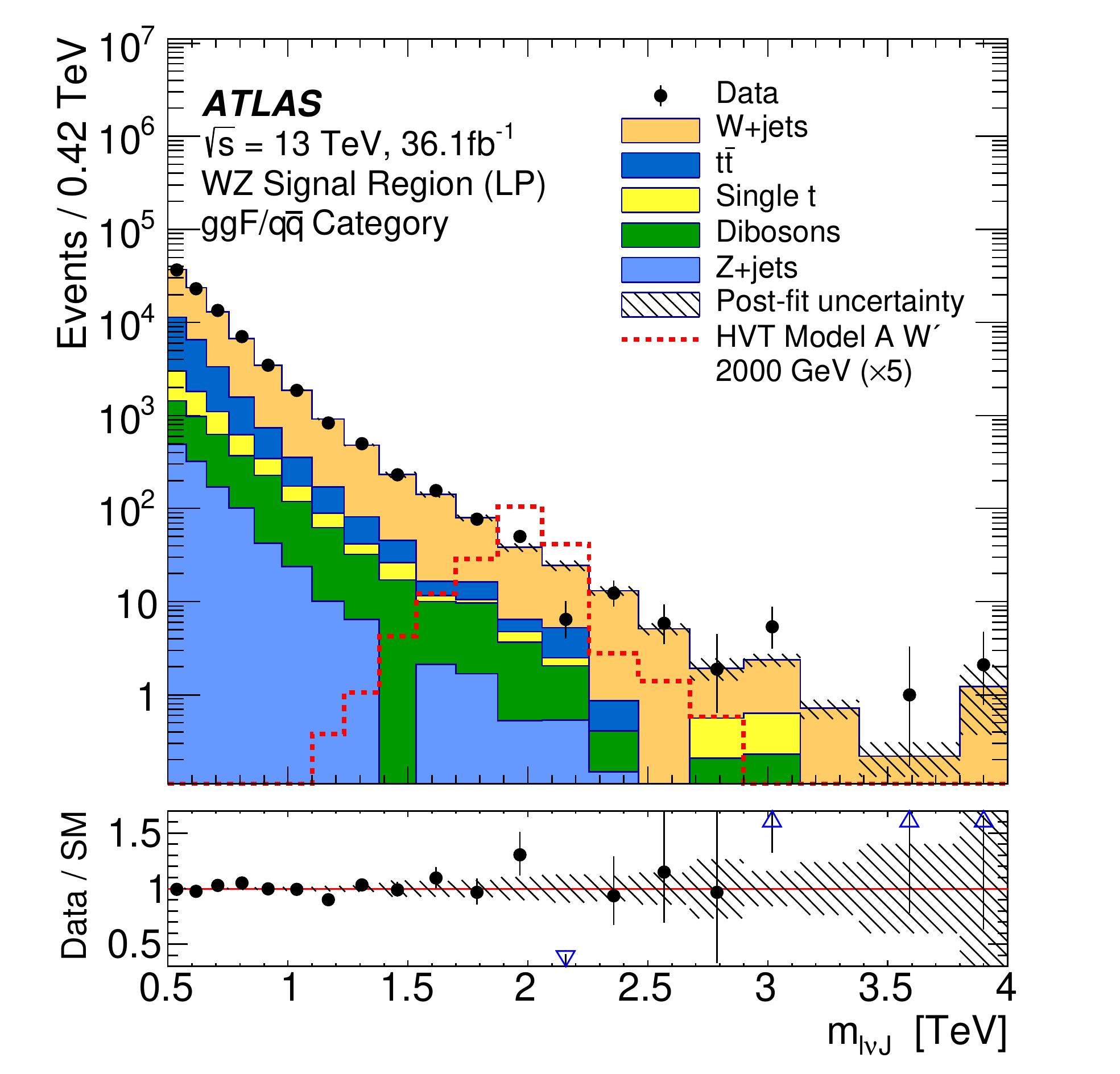}} 
\vspace{-0.4cm}
\subfigure[][\label{fig:pm2a}]{ \includegraphics[width=0.4\hsize]{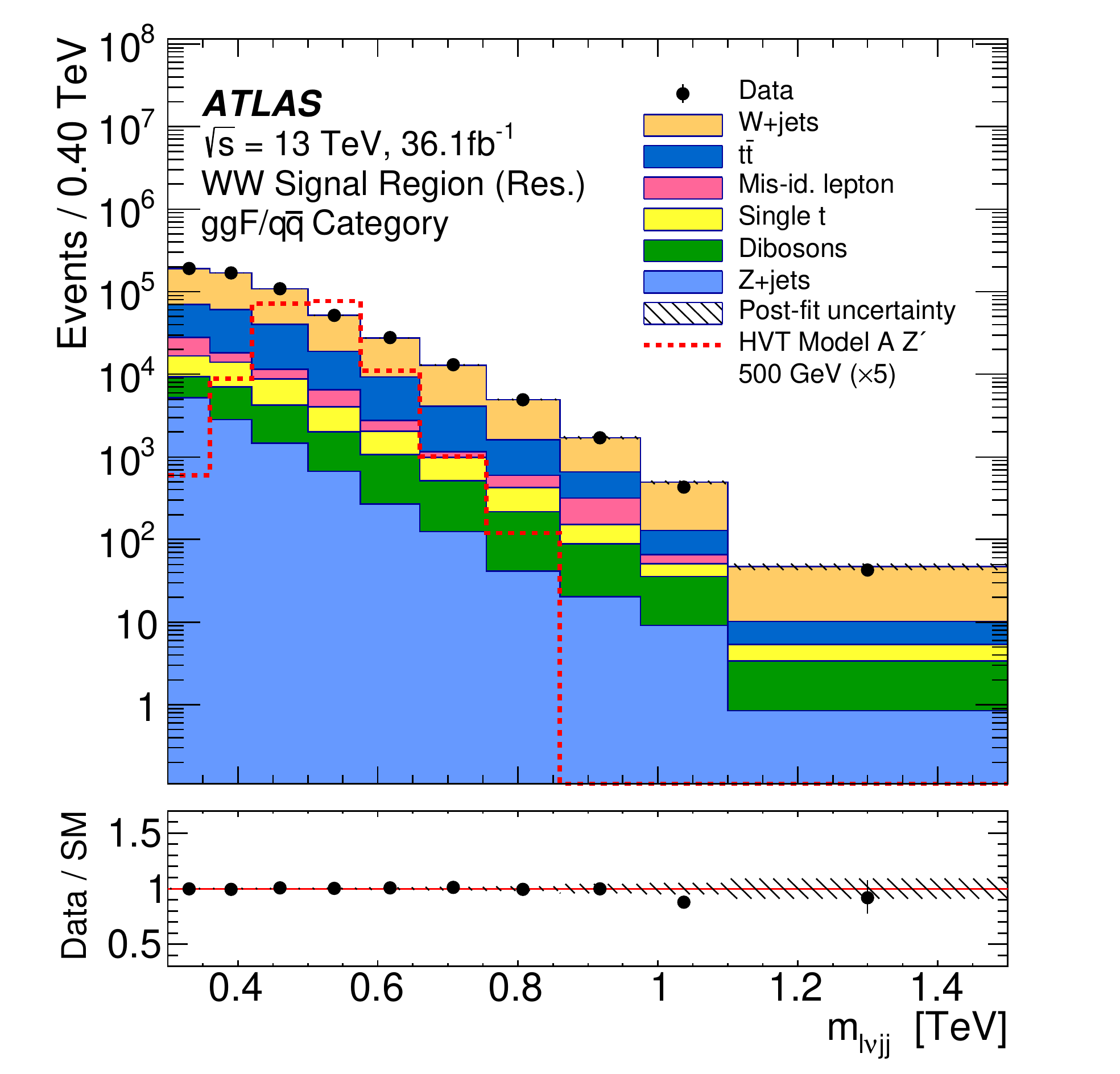}}
\subfigure[][\label{fig:pm2b}]{ \includegraphics[width=0.4\hsize]{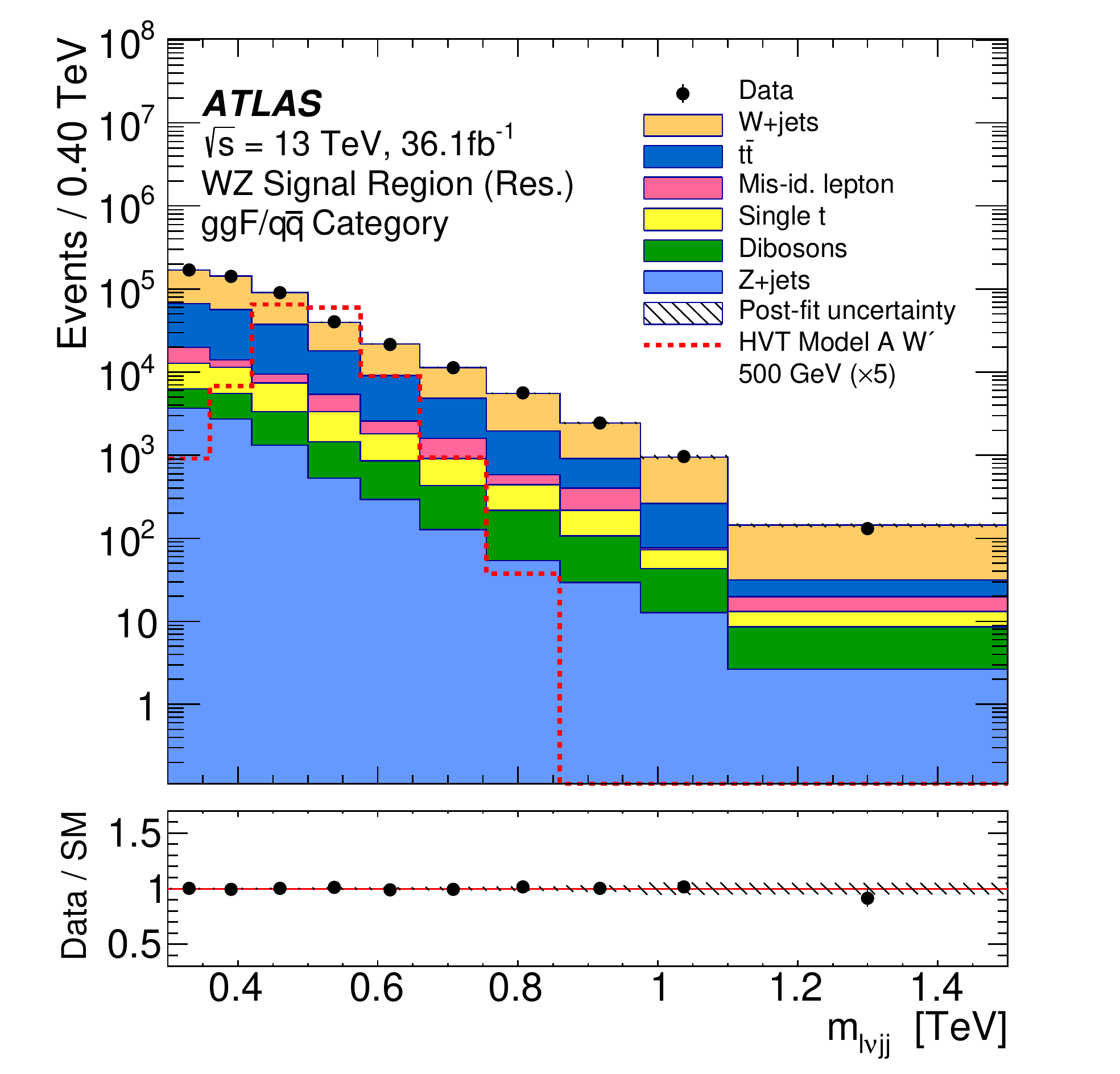}}
\caption{Post-fit signal region $m(WV)$ distributions in the $\mathrm{ggF/q\bar{q}}$ category. The merged high-purity (HP) sample of \subref{fig:pm0a} $WW$ and \subref{fig:pm0b} $WZ$ events, the merged low-purity (LP) sample of \subref{fig:pm1a} $WW$ and \subref{fig:pm1b} $WZ$ events and the resolved (Res.) sample of \subref{fig:pm2a} $WW$ and \subref{fig:pm2b} $WZ$ events are presented. The expected background is shown after the profile likelihood fit to the data, and signal predictions are overlaid. The HVT Model A signal at 2000 \gev\ is presented for the merged analysis, while the 500 \gev\ signal is shown in the resolved topology. The band denotes the statistical and systematic uncertainty in the background after the fit to the data. The lower panels show the ratio of the observed data to the estimated SM background. The distribution of events is shown per mass interval corresponding to the penultimate bin width, while the overflow events are included in the last bin.}
\label{fig:postFit:ggF}
\end{figure}

\begin{figure}[]
\centering
\subfigure[][\label{fig:limitVBF0}]{\includegraphics[width=0.48\linewidth]{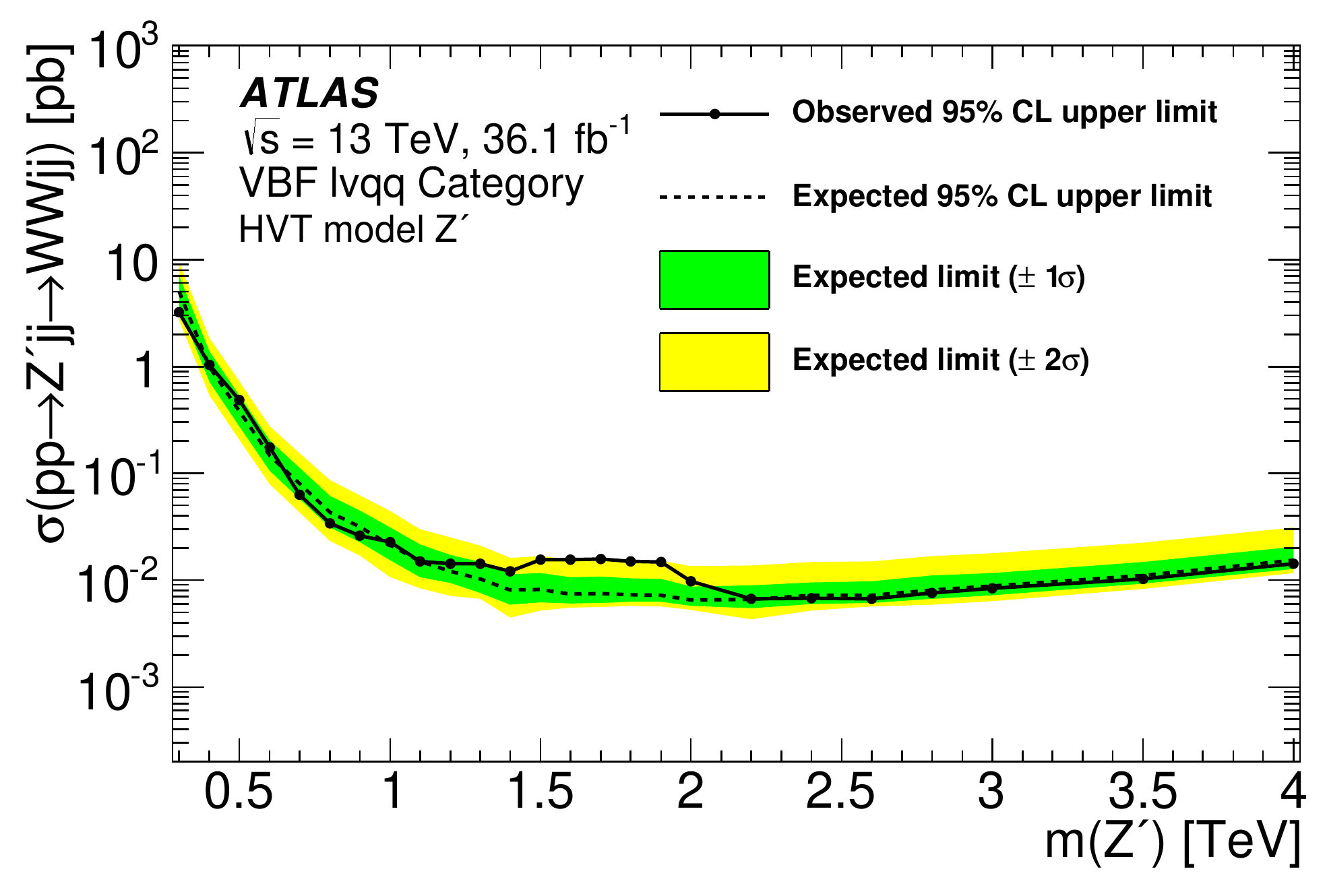}}
\subfigure[][\label{fig:limitVBF1}]{\includegraphics[width=0.48\linewidth]{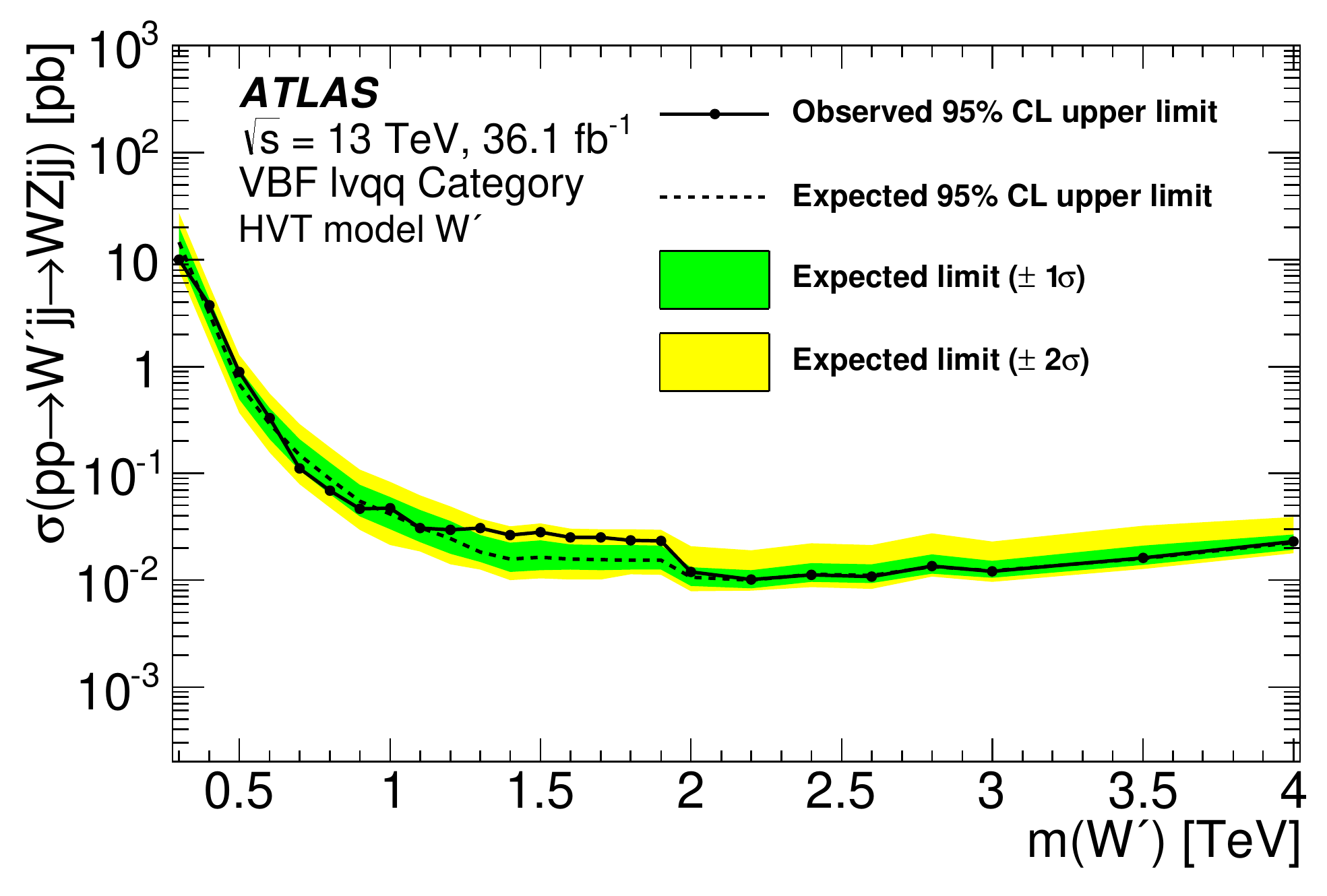}}
\subfigure[][\label{fig:limitVBF2}]{\includegraphics[width=0.48\linewidth]{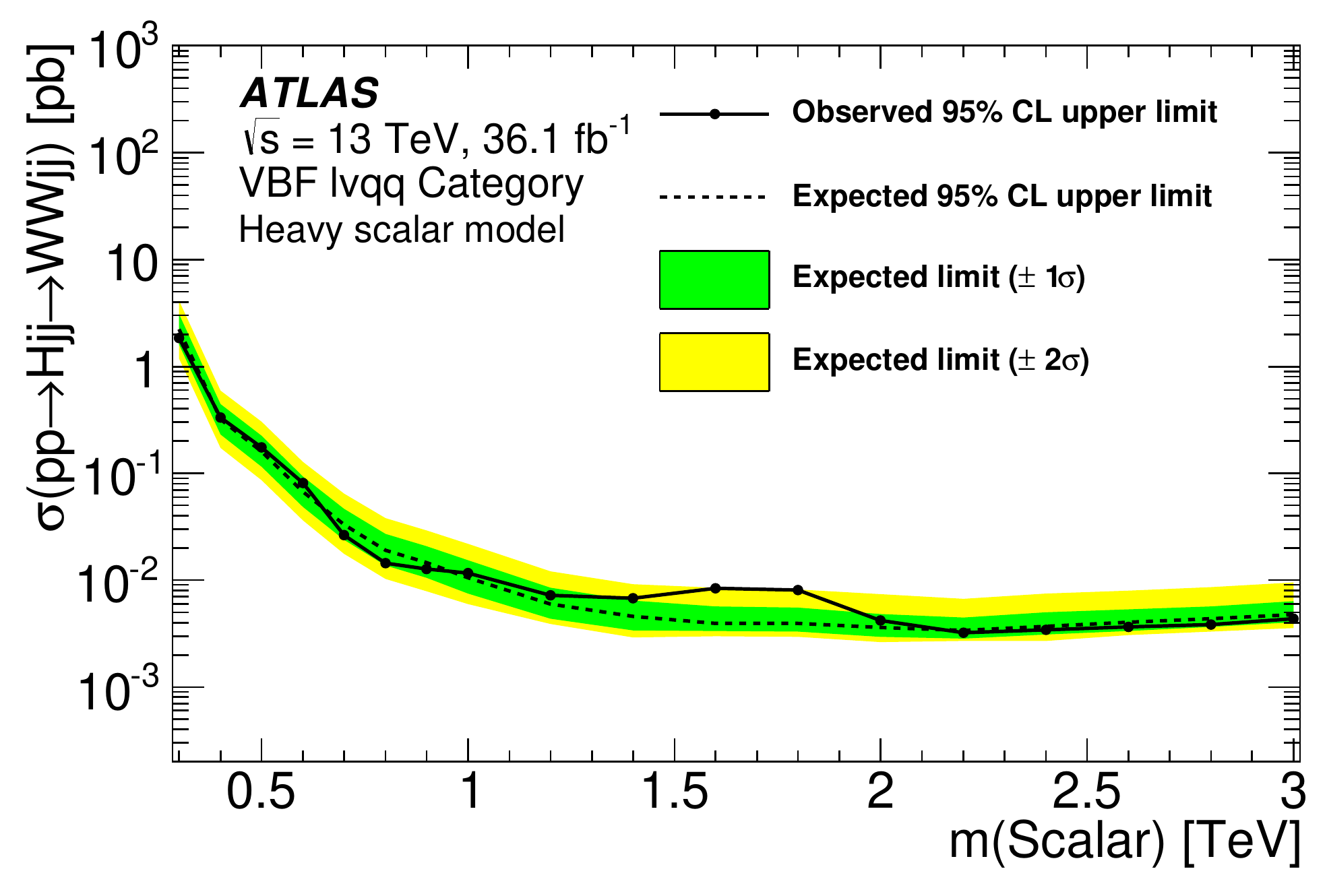}}
\caption{The observed and expected cross-section upper limits at the 95\% confidence level for $WV$ production in the VBF category are presented as a function of the resonance mass. The dots in the observed limit curve represent the generated resonance mass values. Interpretations for \subref{fig:limitVBF0} HVT $Z'$, \subref{fig:limitVBF1} HVT $W'$ and \subref{fig:limitVBF2} heavy scalar signals, $H$, produced via VBF are shown. The mass region greater than 1500 \gev\ is covered by two bins in $m(WV)$. }  
\label{fig:limitVBF}
\end{figure}

\begin{figure}[]
\centering
\subfigure[][\label{fig:limitggF0}]{\includegraphics[width=0.48\linewidth]{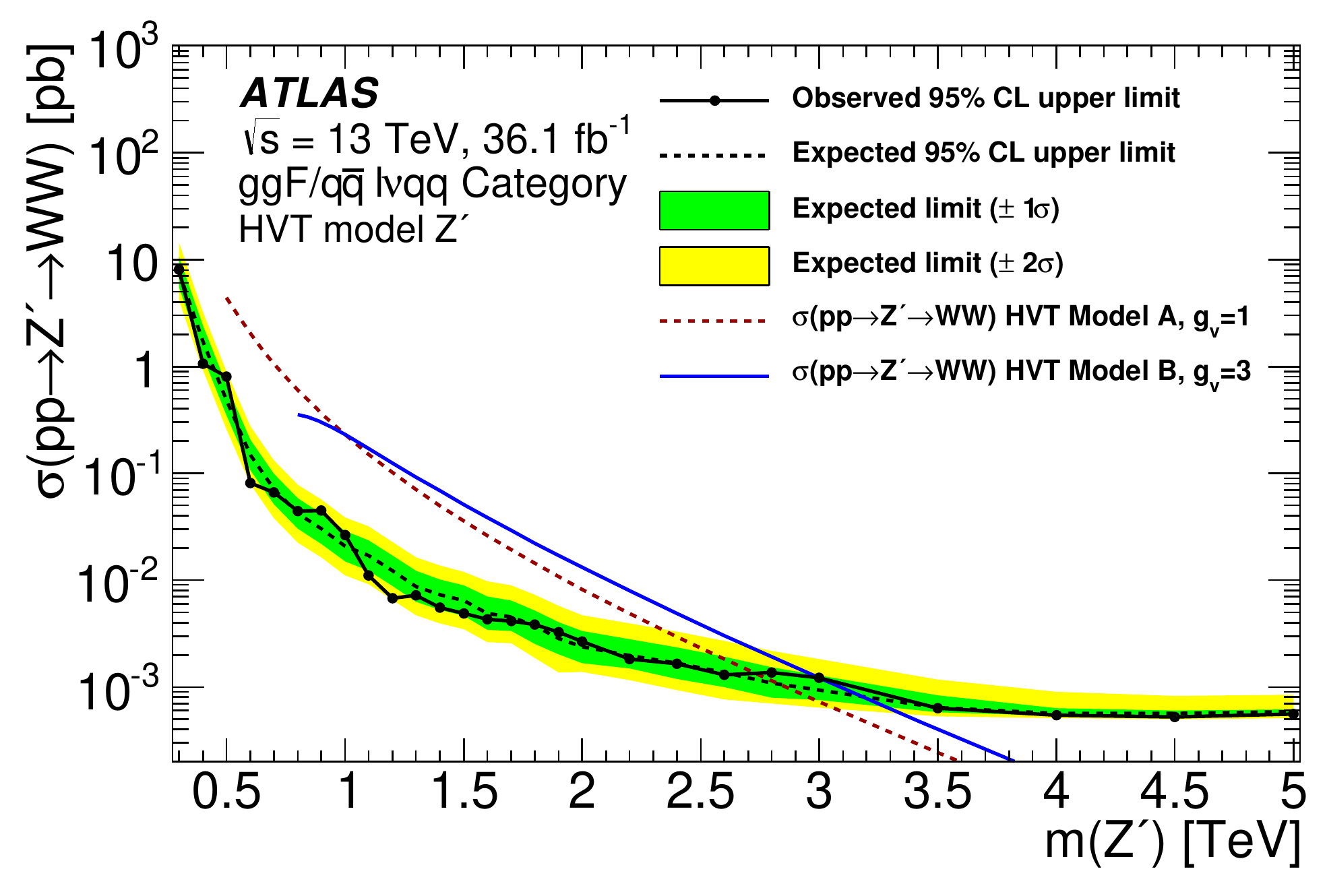}}
\subfigure[][\label{fig:limitggF1}]{\includegraphics[width=0.48\linewidth]{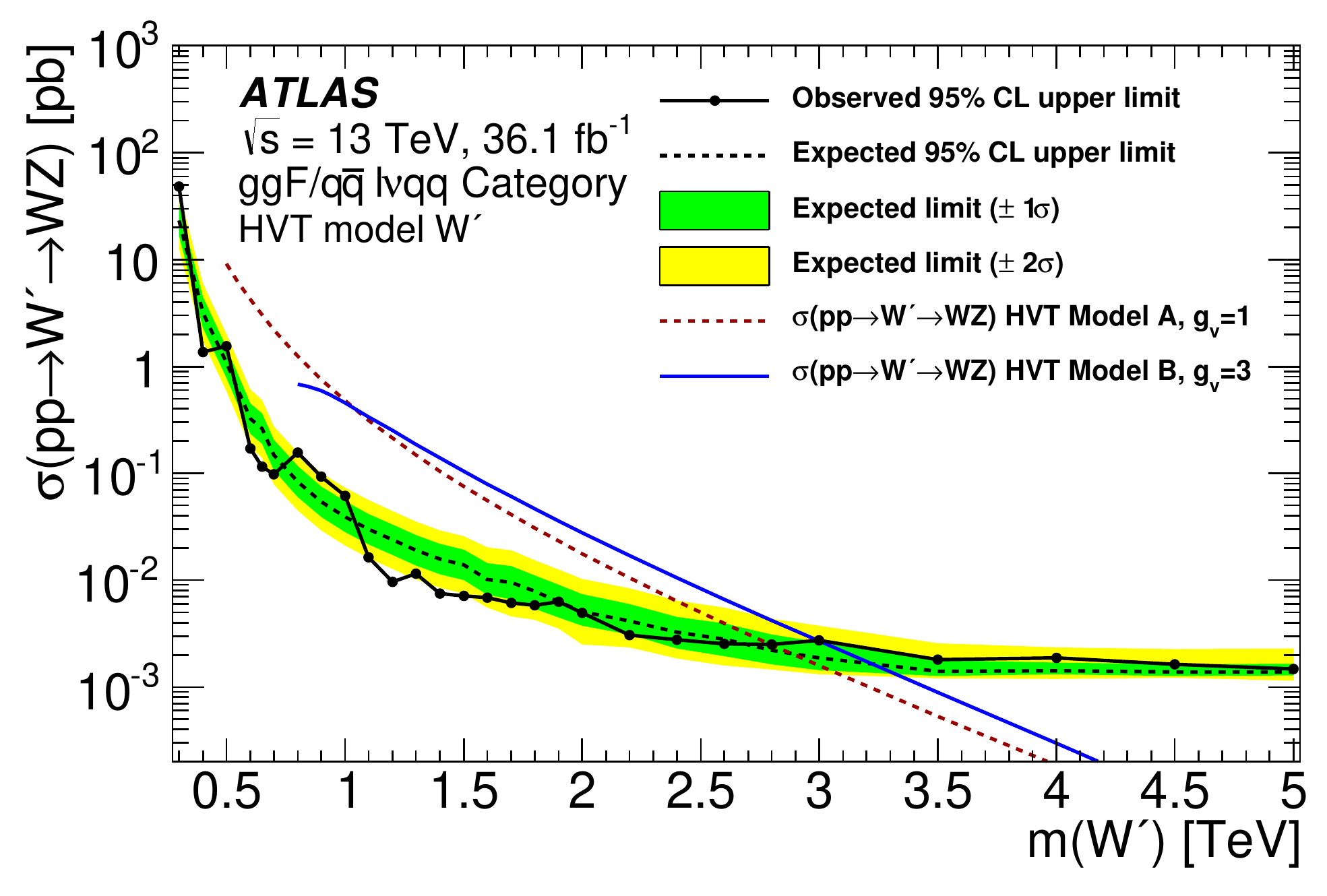}}
\subfigure[][\label{fig:limitggF2}]{\includegraphics[width=0.48\linewidth]{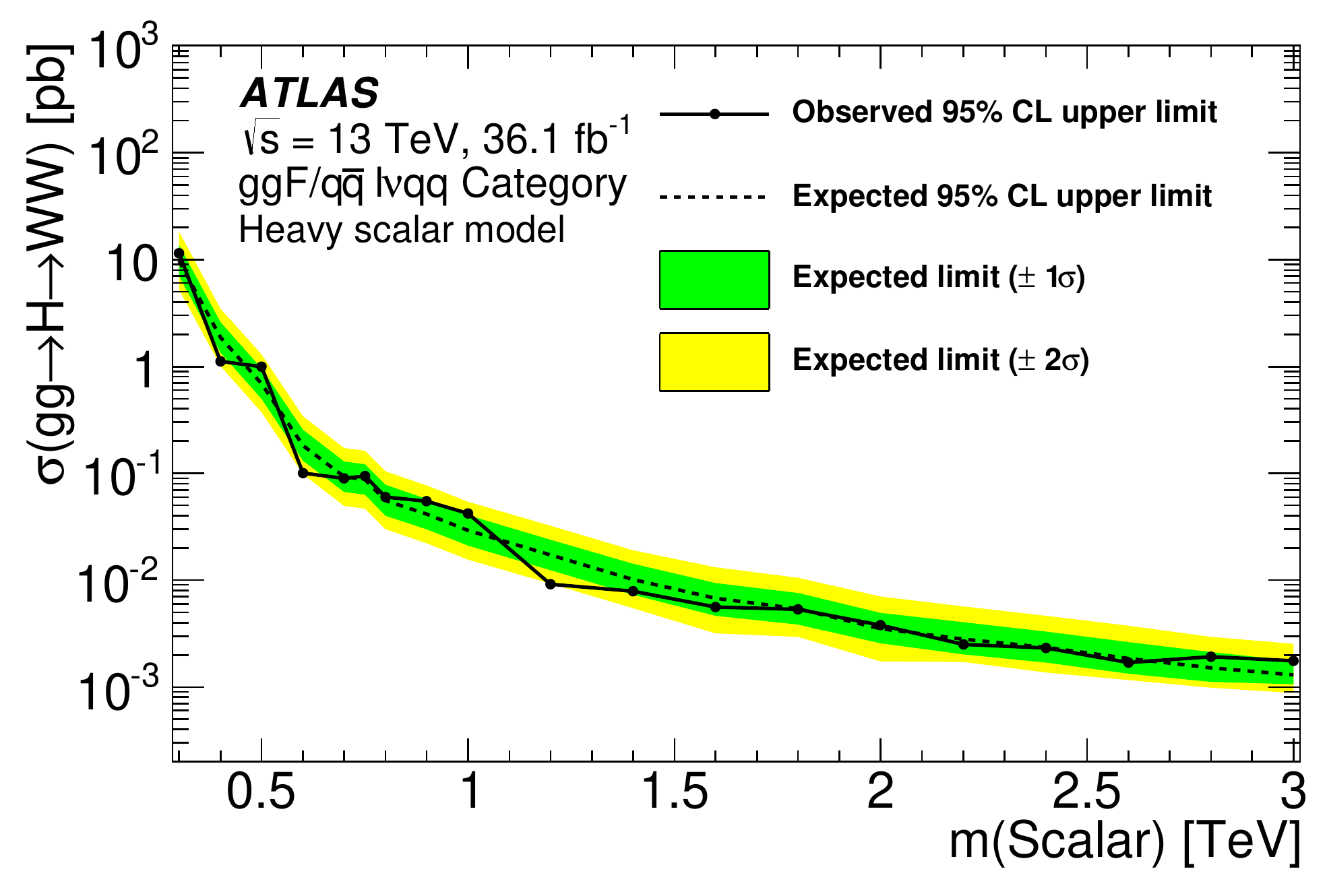}}
\subfigure[][\label{fig:limitggF3}]{\includegraphics[width=0.48\linewidth]{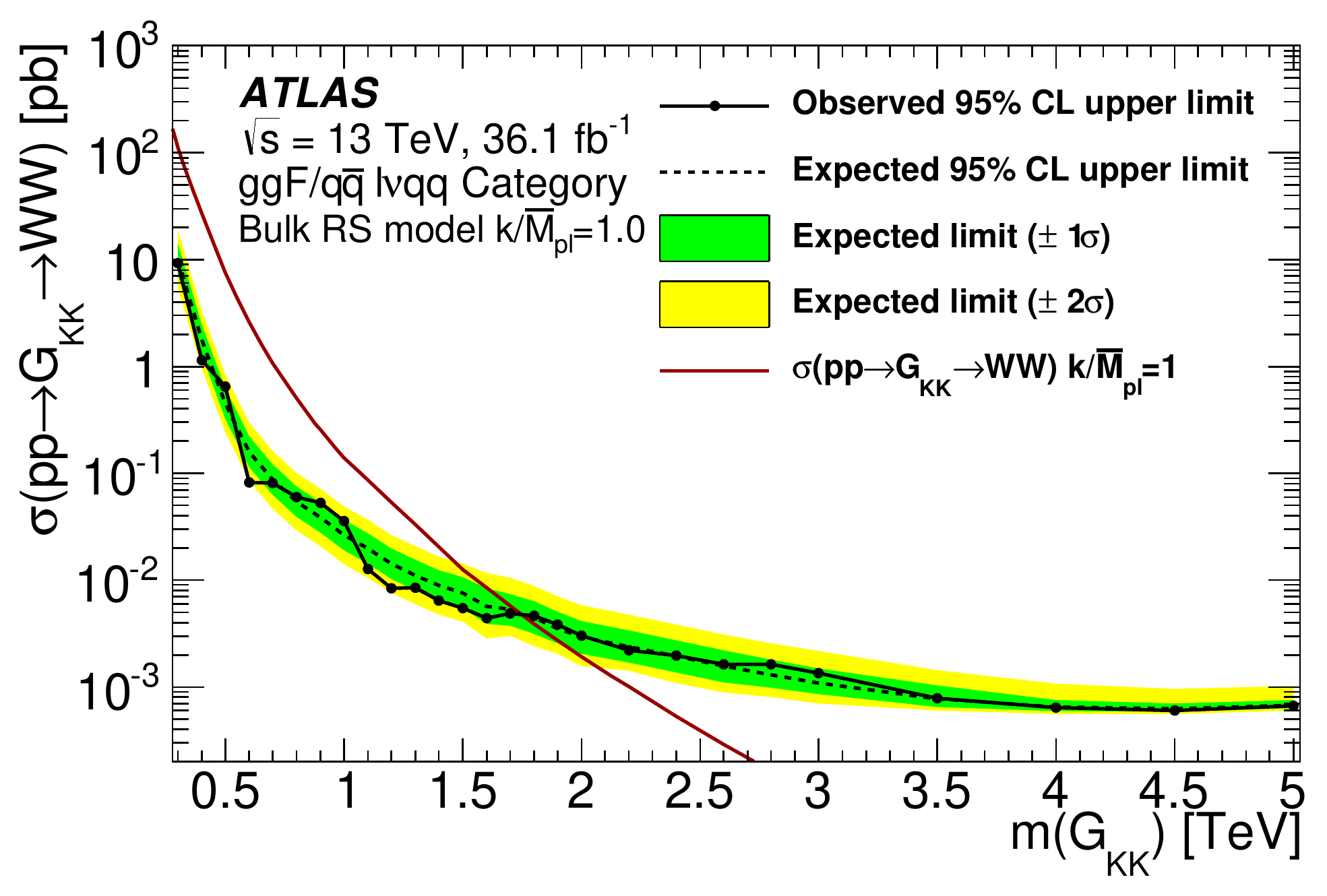}}
\caption{The observed and expected cross-section upper limits at the 95\% confidence level for $WV$ production in the $\mathrm{ggF/q\bar{q}}$ category are presented as a function of the resonance mass. Interpretations for \subref{fig:limitggF0} HVT $WW$, \subref{fig:limitggF0} HVT $WZ$, \subref{fig:limitggF2} scalar $H \to WW$ and \subref{fig:limitggF3} $G_\mathrm{KK}$ produced via gluon--gluon fusion or quark--antiquark annihilation are presented. The red and blue curves, where available, show the predicted signal cross section as a function of resonance mass.}
\label{fig:limitggF}
\end{figure}

\begin{table}[h]
\centering
\caption{Observed and expected excluded masses at the 95\% confidence level for various signal hypotheses as extracted from the $\mathrm{ggF/q\bar{q}}$ category.} 
\label{tab:limres}
\vspace{5pt}
\begin{tabular}{cccc}
\hline
\hline
\multicolumn{4}{c}{$WW$ Selection}\\
\hline
Excluded & \multicolumn{2}{c}{HVT} & RS $G_\mathrm{KK}$ \\       
Masses     &  Model A  & Model B  & $k/\bar{M}_{{\text{Pl}}}$ = 1.0 \\
\hline
Observed  & <2750 \gev & <3000 \gev& <1750 \gev \\
Expected  & <2850 \gev & <3150 \gev& <1750 \gev \\
\hline
\multicolumn{4}{c}{}\\
\hline
\multicolumn{4}{c}{$WZ$ Selection}\\
\hline
Excluded & \multicolumn{2}{c}{HVT} & \\
Masses     &  Model A  & Model B  & \\
\hline
Observed  & <2800 \gev & <3000 \gev& \\
Expected  & <2900 \gev & <3200 \gev&\\
\hline
\hline
\end{tabular}              
\end{table}

%-------------------------------------------------------------------------------
\FloatBarrier
\section{Conclusions}
\label{sec:conclusion}
A search is conducted for resonant $WW$ and $WZ$ production decaying into semileptonic ($\ell \nu qq$) final states using 36.1 \ifb\ of $pp$ collision data collected at a centre-of-mass energy of $\sqrt{s} =$ 13 \tev\ by the ATLAS detector at the LHC during 2015 and 2016. The analysis is carried out in two different kinematic topologies of the hadronically decaying $W/Z$ boson, which can be reconstructed either as two small-$R$ jets or one large-$R$ jet. The data are compatible with the Standard Model background hypothesis and the largest local excess observed is approximately 2.7~$\sigma$, which is not significant. Limits on the production cross section are obtained as a function of the resonance mass for models predicting a narrow scalar boson, a heavy spin-1 vector boson and a spin-2 $\mathrm{KK}$ graviton. Two different production modes are considered, the vector-boson fusion and the gluon--gluon fusion or quark--antiquark annihilation, and independent limits are set. Masses below 2730 \gev\ and 3000 \gev\ are excluded at 95\% CL for the $Z'$ in models A and B of the HVT parametrization, respectively. For the $W'$ resonance, the corresponding limits obtained exclude masses below 2800 \gev\ and 2990 \gev. Additionally, RS $G_{KK}$ signals with $k/\bar{M}_{{\text{Pl}}}$ = 1.0 produced via gluon--gluon fusion are excluded at 95\% CL below 1750 \gev. This search has significantly extended previous ATLAS high-mass limits~\cite{EXOT-2016-01}, by 390-660 \gev, depending on the model.

%-------------------------------------------------------------------------------
\newpage

\section*{Acknowledgements}
\label{sec:acknowledgemnts}
% Acknowledgements for papers with collision data
% Version 16-Nov-2017

% Standard acknowledgements start here
%----------------------------------------------
We thank CERN for the very successful operation of the LHC, as well as the
support staff from our institutions without whom ATLAS could not be
operated efficiently.

We acknowledge the support of ANPCyT, Argentina; YerPhI, Armenia; ARC, Australia; BMWFW and FWF, Austria; ANAS, Azerbaijan; SSTC, Belarus; CNPq and FAPESP, Brazil; NSERC, NRC and CFI, Canada; CERN; CONICYT, Chile; CAS, MOST and NSFC, China; COLCIENCIAS, Colombia; MSMT CR, MPO CR and VSC CR, Czech Republic; DNRF and DNSRC, Denmark; IN2P3-CNRS, CEA-DRF/IRFU, France; SRNSF, Georgia; BMBF, HGF, and MPG, Germany; GSRT, Greece; RGC, Hong Kong SAR, China; ISF, I-CORE and Benoziyo Center, Israel; INFN, Italy; MEXT and JSPS, Japan; CNRST, Morocco; NWO, Netherlands; RCN, Norway; MNiSW and NCN, Poland; FCT, Portugal; MNE/IFA, Romania; MES of Russia and NRC KI, Russian Federation; JINR; MESTD, Serbia; MSSR, Slovakia; ARRS and MIZ\v{S}, Slovenia; DST/NRF, South Africa; MINECO, Spain; SRC and Wallenberg Foundation, Sweden; SERI, SNSF and Cantons of Bern and Geneva, Switzerland; MOST, Taiwan; TAEK, Turkey; STFC, United Kingdom; DOE and NSF, United States of America. In addition, individual groups and members have received support from BCKDF, the Canada Council, CANARIE, CRC, Compute Canada, FQRNT, and the Ontario Innovation Trust, Canada; EPLANET, ERC, ERDF, FP7, Horizon 2020 and Marie Sk{\l}odowska-Curie Actions, European Union; Investissements d'Avenir Labex and Idex, ANR, R{\'e}gion Auvergne and Fondation Partager le Savoir, France; DFG and AvH Foundation, Germany; Herakleitos, Thales and Aristeia programmes co-financed by EU-ESF and the Greek NSRF; BSF, GIF and Minerva, Israel; BRF, Norway; CERCA Programme Generalitat de Catalunya, Generalitat Valenciana, Spain; the Royal Society and Leverhulme Trust, United Kingdom.

The crucial computing support from all WLCG partners is acknowledged gratefully, in particular from CERN, the ATLAS Tier-1 facilities at TRIUMF (Canada), NDGF (Denmark, Norway, Sweden), CC-IN2P3 (France), KIT/GridKA (Germany), INFN-CNAF (Italy), NL-T1 (Netherlands), PIC (Spain), ASGC (Taiwan), RAL (UK) and BNL (USA), the Tier-2 facilities worldwide and large non-WLCG resource providers. Major contributors of computing resources are listed in Ref.~\cite{ATL-GEN-PUB-2016-002}.
%----------------------------------------------

%-------------------------------------------------------------------------------
\printbibliography
%-------------------------------------------------------------------------------

\newpage 
% ATLAS Collaboration author list
% Data extracted on 02-Sep-2017 for paper reference EXOT-2016-28
%\documentclass[11pt]{article}
%\usepackage{a4wide}\begin{document}
\begin{flushleft}
{\Large The ATLAS Collaboration}

\bigskip

M.~Aaboud$^\textrm{\scriptsize 137d}$,
G.~Aad$^\textrm{\scriptsize 88}$,
B.~Abbott$^\textrm{\scriptsize 115}$,
O.~Abdinov$^\textrm{\scriptsize 12}$$^{,*}$,
B.~Abeloos$^\textrm{\scriptsize 119}$,
S.H.~Abidi$^\textrm{\scriptsize 161}$,
O.S.~AbouZeid$^\textrm{\scriptsize 139}$,
N.L.~Abraham$^\textrm{\scriptsize 151}$,
H.~Abramowicz$^\textrm{\scriptsize 155}$,
H.~Abreu$^\textrm{\scriptsize 154}$
% ...
\bigskip
\\
% ...
$^{12}$ Institute of Physics, Azerbaijan Academy of Sciences, Baku, Azerbaijan\\
% ...
$^{88}$ CPPM, Aix-Marseille Universit{\'e} and CNRS/IN2P3, Marseille, France\\
% ...
$^{115}$ Homer L. Dodge Department of Physics and Astronomy, University of Oklahoma, Norman OK, United States of America\\
% ...
$^{119}$ LAL, Univ. Paris-Sud, CNRS/IN2P3, Universit{\'e} Paris-Saclay, Orsay, France\\
% ...
$^{137}$ $^{(a)}$ Facult{\'e} des Sciences Ain Chock, R{\'e}seau Universitaire de Physique des Hautes Energies - Universit{\'e} Hassan II, Casablanca; $^{(b)}$ Centre National de l'Energie des Sciences Techniques Nucleaires, Rabat; $^{(c)}$ Facult{\'e} des Sciences Semlalia, Universit{\'e} Cadi Ayyad, LPHEA-Marrakech; $^{(d)}$ Facult{\'e} des Sciences, Universit{\'e} Mohamed Premier and LPTPM, Oujda; $^{(e)}$ Facult{\'e} des sciences, Universit{\'e} Mohammed V, Rabat, Morocco\\
% ...
$^{139}$ Santa Cruz Institute for Particle Physics, University of California Santa Cruz, Santa Cruz CA, United States of America\\
% ...
$^{151}$ Department of Physics and Astronomy, University of Sussex, Brighton, United Kingdom\\
% ...
$^{154}$ Department of Physics, Technion: Israel Institute of Technology, Haifa, Israel\\
$^{155}$ Raymond and Beverly Sackler School of Physics and Astronomy, Tel Aviv University, Tel Aviv, Israel\\
% ...
$^{161}$ Department of Physics, University of Toronto, Toronto ON, Canada\\
% ...
$^{a}$ Also at Department of Physics, King's College London, London, United Kingdom\\
$^{b}$ Also at Institute of Physics, Azerbaijan Academy of Sciences, Baku, Azerbaijan\\
$^{c}$ Also at Novosibirsk State University, Novosibirsk, Russia\\
$^{d}$ Also at TRIUMF, Vancouver BC, Canada\\
$^{e}$ Also at Department of Physics {\&} Astronomy, University of Louisville, Louisville, KY, United States of America\\
% ...
$^{*}$ Deceased
\end{flushleft}

%\end{document}
% Created with xml2latex.py

%-------------------------------------------------------------------------------
%\clearpage
%\part*{Auxiliary material}
%\addcontentsline{toc}{part}{Auxiliary material}
%\clearpage
%\appendix
%\input{texfiles/appendix.tex}         
%-------------------------------------------------------------------------------

\end{document}